\documentclass[11pt]{article}

\usepackage{graphicx}
\usepackage{amsmath}
\usepackage{amsfonts}
\usepackage{fancyhdr} 

\usepackage{wrapfig}

\usepackage{bbold}

\usepackage{float}

\newtheorem{lem}{Lemma}
\newtheorem{thm}{Theorem}

\newtheorem{prop}[lem]{Proposition}
\newtheorem{cor}[lem]{Corollary}
\newtheorem{defn}{Definition}
\newtheorem{example}[lem]{Example}
\newtheorem{notation}[defn]{Notation}
\newtheorem{alg}{Algorithm}

\numberwithin{equation}{section}
\numberwithin{thm}{section}
\numberwithin{lem}{section}
\numberwithin{defn}{section}
\numberwithin{alg}{section}

\newcommand{\sss}{\mathsf{S}}
\newcommand{\supp}{\mbox{supp}\,}
\newcommand{\real}{\mathbb{R}}
\newcommand{\integer}{\mathbb{Z}}

\newenvironment{pf}{\noindent {\em Proof}.\ \ }{\hspace*{\fill}\rule{.5ex}{1.4ex}\,}

\newcommand{\nat}{\mathbb{N}}

\newcommand{\gtr}{G^{(\tau,R)}}

\newcommand{\LtM}{\mathfrak{L}^{\tau}_M}
\newcommand{\Ltns}{\mathfrak{L}^{\tau^{(n)}}_{n,s}}
\newcommand{\Ltn}{\mathfrak{L}^{\tau^{(n)}}_{n,\sigma_d}}

\newcommand{\mpp}{\mathsf{p}}

\title{The purely singular 1-D acoustic reflection problem}
\author{Peter C.~Gibson \footnote{Dept.~of Mathematics \& Statistics, York University, 4700 Keele St., Toronto, Ontario, Canada, M3J~1P3, $\mathtt{pcgibson@yorku.ca}$} \footnote{Research supported by NSERC and MITACS} }

\begin{document}

\maketitle

\begin{abstract} 
This paper analyzes the nonlinear correspondence between the reflectivity profile (model) and the plane wave impulse response at the boundary (data) for a three-dimensional half space consisting of a sequence of homogeneous horizontal layers.  This correspondence is of importance in geophysical imaging, where it has been studied for more than half a century from a variety of perspectives.   The main contribution of the present paper is to derive something new in the context of a time-limited deterministic approach: (i) an exact finite (non-asymptotic) formula for the data in terms of the model, (ii) a corresponding exact inverse algorithm, and (iii) a precise characterization of the inherent nonlinearity.   Regarding (iii), for generic models the correspondence is characterized as a pair of maps, one of which is locally linear, and the other of which is locally polynomial.  Both are determined by a local combinatorial invariant, an integer matrix.  Concerning (ii), the basic inverse algorithm is modified to allow for erroneous amplitude data, taking advantage of the overdeterminacy of the inverse problem to recover the exact model even in cases where the data is badly distorted.   The results are illustrated with numerical examples.
\end{abstract}

\newpage

\tableofcontents

\newpage

\section{Introduction}

\pagestyle{fancyplain}

The acoustic reflection problem in the setting of a layered three-dimensional half space is fundamental to seismic imaging,  a connection in which it has been studied for more than half a century.   Two recent developments motivate taking a fresh look at this problem, pared down to its simplest form.   Firstly, recent progress in superresolution \cite{Ca:2012} opens up the possibility of working with the true impulse response.  And secondly, there is an increasing effort among seismologist to incorporate non-linear effects such as multiple reflections into the data processing flow \cite{MaUrDeHo:2009}\cite{In:2009}\cite{GrVe:2011}.   Intuitively, multiple reflections are a highly redundant source of potential information about material properties. But the redundancy is wasted without a precise elucidation of how material properties are encoded---hence the development of multiple suppression (or elimination) as a standard seismic signal processing technique \cite[Chapter~2]{Yi:2001}.  Indeed, as pointed out the recent survey \cite{Sy:2009}, seismic signal processing is largely predicated on linearization and single scattering.  The present paper aims to analyze fully the non-linear relationship between material properties and boundary measurements for a particular formulation of the forward and inverse problems pertinent to a layered half space---including a comprehensive treatment of multiple reflections.  The goal is to establish a mathematically rigorous theory that gives a clear perspective on the deterministic approach.  Several new results are established, including:
\begin{itemize}
\item exact polynomial formulas for the impulse response (referred to as data) in terms of physical parameters (referred to collectively as the model);
\item the existence of a local combinatorial invariant, an integer matrix, that facilitates a precise characterization of the forward and inverse mappings between model and data;
\item a fast exact inverse algorithm that avoids downward continuation and that exploits the inherent redundancy of the data to correct for erroneous amplitudes.
\end{itemize}

The remainder of the introduction is divided into sections as follows.  Section~\ref{sec-formulation} formulates precisely the forward and inverse problems that are the subject of the rest of the paper, and compares the given formulation to earlier treatments.  Section~\ref{sec-results-overview} summarizes the paper's main results.    And Section~\ref{sec-literature} cites some additional related literature.

\subsection{Problem statement\label{sec-formulation}}

\begin{figure}[h]
\hspace*{.5in}\parbox{4.25in}{
\fbox{\includegraphics[width=4.25in]{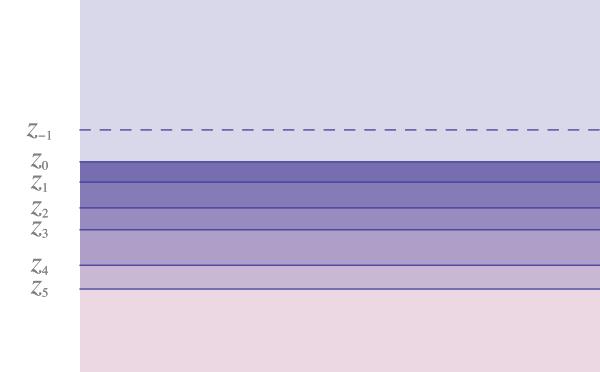}}
\caption{ A layered medium, with $z$ increasing downward.}\label{fig-layered}
}
\end{figure}

The physical setup is as follows.   Consider a three-dimensional acoustic medium with coordinates $(x,y,z)$ such that the medium varies only in the $z$-direction and such that the density $\rho(z)$ and bulk modulus $K(z)$ are piecewise constant with respect to $z$, having jumps at points
\[
z_0<z_1<\cdots<z_M,
\]
where $M\geq 1$.  Let $z_{-1}<z_0$ denote a fixed reference depth in the homogeneous half space $z<z_0$ from which signals in the form of traveling plane waves may be transmitted and received.  In keeping with the geophysical perspective, the $z$-coordinate will be interpreted as depth, and it will be depicted as increasing downward, as in Figure~\ref{fig-layered}. 
The depth range $z_{n-1}<z<z_n$ with be referred to as the $n$th layer, for $1\leq n\leq M$.   Thus there are $M$ layers and $M+1$ interfaces, the latter being located at depths $z_0,z_1,\ldots,z_M$.    

Let $u(t,z)$ denote the velocity (in the $z$-direction) of a material particle at depth $z$ and time $t$, and let $p(t,z)$ denote the pressure.   The medium evolves according to the coupled one-dimensional equations
\begin{subequations}
\begin{align}
\rho\frac{\partial u}{\partial t}+\frac{\partial p}{\partial z}&=0\label{wave1}\\
\frac{1}{K}\frac{\partial p}{\partial t}+\frac{\partial u}{\partial z}&=0.\label{wave2}
\end{align}
\end{subequations}
In analyzing the solution to the above system, there is a choice to be made between working with $u(t,z)$ or $p(t,z)$ (or a certain combination or the two).  The situation is essentially the same whatever the choice; for the sake of definiteness the present article will focus on the velocity field $u(t,z)$, which is the quantity measured by a coil/magnet geophone, for instance.
The initial conditions corresponding to a plane wave unit impulse propagating downward from $z_{-1}$ are
\begin{equation}\label{initial}
\begin{split}
u(0,z)&=\delta(z-z_{-1})\\
 p(0,z)&=\sqrt{K(z_{-1})\rho(z_{-1})}\;\delta(z-z_{-1}).
 \end{split}
\end{equation}
Let $G(t)$ denote the (velocity) impulse response at $z_{-1}$, so that 
\begin{equation}\label{response}
G(t)=u(t,z_{-1})\quad\quad(t>0),
\end{equation}
the solution at depth $z_{-1}$ to the system (\ref{wave1},\ref{wave2},\ref{initial}).   

The work \cite[Chapter~3]{FoGaPaSo:2007} of Fouque \emph{et al.}\ summarizes very clearly the standard theory concerning propogation of waves in a piecewise constant layered medium, including a derivation of the governing equations (\ref{wave1},\ref{wave2}) from physical principles; it will serve as a principal reference.  See \cite[Section~2]{BuBu:1983} for an alternate treatment of the same material.  The following facts, proved in  \cite[Chapter~3]{FoGaPaSo:2007} and elsewhere, serve as a starting point for the present paper.   For $1\leq n\leq M$, let $\tau_n$ denote the two-way travel time (for a traveling wave) across the $n$th layer of the above $M$-layer medium, and let $\tau_0$ denote the two-way travel time from depth $z_{-1}$ to $z_0$.   For $0\leq n\leq M$, let $R_n$ denote the reflection coefficient at depth $z_n$ relative to a wave traveling toward the interface from above.  Letting $K_n$ and $\rho_n$ denote the density and bulk modulus inside the $n$th layer---with $K_{-1},\rho_{-1}$ and $K_{M+1},\rho_{M+1}$ 
denoting the respective values at $z_{-1}$ and any point $z_{M+1}$ below $z_{M}$---the travel times and reflectivities are given by the formulas
\begin{equation}\label{tau-R-formulas}
\tau_n=\frac{2(z_n-z_{n-1})}{\sqrt{K_n/\rho_n}}\quad\mbox{ and }\quad R_n=\frac{\sqrt{K_n\rho_n}-\sqrt{K_{n+1}\rho_{n+1}}}{\sqrt{K_n\rho_n}+\sqrt{K_{n+1}\rho_{n+1}}},
\end{equation}
for $0\leq n\leq M$.  Note that $-1<R_n<1$ by virtue of (\ref{tau-R-formulas}).  Let $(\tau,R)$ denote the pair of sequences
\[
\tau=(\tau_0,\ldots,\tau_M)\quad\mbox{ and }\quad R=(R_0,\ldots,R_M),
\]
and let $|\tau|$ denote the total two-way travel time from $z_{-1}$ to $z_M$,
\[
|\tau|=\tau_0+\cdots+\tau_M.
\]
The pair $(\tau,R)$ will be referred to as a model; it completely determines $G(t)$ and hence represents the underlying physical structure insofar as concerns the impulse response.   To emphasize this dependence, the impulse response as defined in (\ref{response}) will be denoted $\gtr(t)$.   Moreover, the impulse response has the structure of a delta train of the form
\begin{equation}\label{normalform1}
G^{(\tau,R)}(t)=\sum_{n=1}^\infty\alpha_n\delta(t-\sigma_n).   
\end{equation}
Without loss of generality the representation (\ref{normalform1}) will be taken to be in normal form, whereby each $\alpha_n\neq0$ and the arrival times $\sigma_n$ are in their natural order,
\[
\sigma_1<\sigma_2<\sigma_2<\cdots.
\]
In practice the impulse response (\ref{normalform1}) can only be known for some finite time interval $[0,t_{\rm max}]$.  If $t_{\rm max}<|\tau|$, say
 \[
 \tau_0+\cdots+\tau_N\leq t_{\rm max}< \tau_0+\cdots+\tau_{N+1},
 \]
 then the terms $\tau_{N+1},\ldots,\tau_M$ and $R_{N+1},\ldots,R_M$ have no influence on $\gtr(t)$ for $t\in[0,t_{\rm max}]$.  That is, letting $\tau^\prime=(\tau_0,\ldots,\tau_N)$ and $R^\prime=(R_0,\ldots,R_N)$, 
 \begin{equation}\nonumber
 \gtr(t)=G^{(\tau^\prime,R^\prime)}(t)\mbox{ for } t<t_{\rm max}.
 \end{equation}
 Thus one may as well study just $G^{(\tau^\prime,R^\prime)}$.   Based on this reasoning, let $\chi_{[0,|\tau|]}$ denote the characteristic function of the finite interval $[0,|\tau|]$ and consider the partial impulse response
 \begin{equation}\label{normalrestriction}
 \chi_{[0,|\tau|]}\gtr(t)=\sum_{n=1}^d\alpha_n\delta(t-\sigma_n).
 \end{equation}
Letting $(\sigma,\alpha)$ denote the finite sequences 
 \[
 \sigma=(\sigma_1,\ldots,\sigma_d)\mbox{ and }\alpha=(\alpha_1,\ldots,\alpha_d),
 \]
the central problem of the present paper is to analyze the mapping 
\[
(\tau,R)\mapsto(\sigma,\alpha).
\]
The data $(\sigma,\alpha)$ corresponding to (\ref{normalrestriction}) is well-defined provided that the right-hand side is in normal form, a convention that will be in force from now on.\\[5pt]   
\setlength{\fboxsep}{8pt}
\noindent\fbox{
\parbox{.93\textwidth}{
\textbf{Forward problem}.\ \ Given an $M$-layer model $(\tau,R)$, for some $M\geq 1$, determine the data $(\sigma,\alpha)$ for which equation (\ref{normalrestriction}) holds.\\[10pt]
\textbf{Inverse problem}.\ \ Given the data $(\sigma,\alpha)$ such that equation (\ref{normalrestriction}) holds for an $M$-layer model $(\tau,R)$, for some $M\geq 1$, determine the model.  
}
}\\[5pt]
In the above formulation, the vectors $\tau$ and $R$ are arbitrary, in the sense that any 
\[
\tau\in\real^{M+1}_{>0}\quad\mbox{ and }\quad R\in(-1,1)^{M+1}
\]
comprise a possible model, for any $M\geq1$.   Three features distinguish the above formulation, setting it apart from earlier treatments of the problem:
\begin{enumerate}
\item the data is a delta train and hence purely singular;
\item the $\tau_n$ are arbitrary---they need not all be equal;
\item the data is restricted to finite time.  
\end{enumerate}
The purely singular structure fits the scenario of Cand\`{e}s and Fernandez-Granda's work \cite{Ca:2012} on superresolution.   Their results suggest that, within certain constraints, it may be possible to extract the function
$ \chi_{[0,|\tau|]}\gtr(t)$ from measured data, which one expects to be a convolution of the form  
\[
 \bigl(\gtr\ast f\bigr)(t),
 \]
 for some non-sigular wavelet $f(t)$.   Deconvolution, which has its own long history, will not be considered in the present paper.   While the purely singular structure is suited to superresolution, it is unsuited to classical methods such as that of Gelfand-Levitan, as pointed out by Bube and Burridge \cite[Section~1.2]{BuBu:1983}.  The second item above, that the $\tau_n$ need not be equal, distinguishes the present formulation from several influential papers on the 1-D reflection problem, including that of Berryman-Goupillaud-Waters \cite{BeGoWa:1958}, Kunetz \cite{Ku:1963} and Bube-Burridge \cite{BuBu:1983}.    Each of the latter assumes constant travel times across layers, which turns out to change the problem substantially from the generic case in which travel times across layers are unrelated to one another.   (See below for more on this point.)    Lastly, the finite time restriction of item (3) makes frequency domain methods unsuitable if one is interested in exact formulas rather than series approximations.  This motivates Browning's work \cite{Br:2000} on extrapolation, for example.   Indeed series approximations underpin a range of approaches to the reflection problem, in both 1-D and higher.  See the survey \cite{WeArFe:2003} for an overview based on the Lippmann-Scwinger equation, and \cite{In:2009} for more recent work along the same lines.   The present paper avoids series approximations altogether.

\subsection{Overview of main results\label{sec-results-overview}}

The forward and inverse problems stated in Section~\ref{sec-formulation}  are solved completely, facilitating a precise description of the model-data correspondence.  The picture that emerges will be sketched below, following some remarks on the forward problem.   An explicit solution to the forward problem is stated in Theorem~\ref{thm-amplitude} and its corollories.   The proof of the theorem---deferred to Section~\ref{sec-proof}---entails a comprehesive analysis of scattering sequences, carried out by means of combinatorial arguments tailored specifically to the task.   The resulting formulas have a very nice structure: the amplitudes $\alpha_n$ in the data turn out to be polynomial functions of the reflectivity $R$.  These polynomials are homogeneous, have integer coefficients, and are consequently easy to code up.   The arrival times $\sigma_n$ have an even simpler structure.  Thus the solution to the forward problem leads to a straightforward exact algorithm, Algorithm~\ref{alg-forward}.    

It is proved in Section~\ref{sec-ill-posed} that the mapping from models to data is not globally injective; distinct models of differing dimensions can produce the same data as one another.  This ill-posedness is shown to to stem from alignment of the travel time vector $\tau=(\tau_0,\ldots,\tau_M)$ with the integer lattice $\integer^{M+1}$, such as when the $\tau_n$ are all equal---a basic assumption of numerous papers on the 1-D problem!    The ill-posedness is remedied by discarding a small set of models (having measure zero in each dimension).    As shown in Corollary~\ref{cor-injective}, the mapping from models to data is injective on the remainder, which are referred to as generic models.   

More than this, the set of $M$-layer generic models is naturally partitioned into open sets $X_\psi$.   On each set, the mapping $(\tau,R)\mapsto(\sigma,\alpha)$ is given by a fixed formula, which has the following form.  Viewing $\tau$ and $\sigma$ as row vectors, there is an integer matrix $A_\psi$ (Definition~\ref{defn-enumeration-matrix}) such that 
\begin{equation}\label{arrival-times}
\sigma=\tau A_\psi
\end{equation}
 (valid throughout $X_\psi$).    And each amplitude $\alpha_n$ has the form $p_{\psi,n}(R)$ (also valid throughout $X_\psi$), where $p_{\psi,n}$ is an explicitly given multivariate polynomial determined by $\psi$.   Thus within each $X_\psi$ the mapping decouples into two independant maps $\tau\mapsto\sigma$ and $R\mapsto\alpha$.  The integer matrix $A_\psi$ and the polynomials $p_{\psi,n}$ change from one open set $X_\psi$ to another.   In fact, the integer matrix $A_\psi$ is an invariant that uniquely labels $X_\psi$ and determines the associated polynomials $p_{\psi,n}$.   
The mapping from generic models to data is a diffeomorphism on each $X_\psi$, and it is globally injective when $X_\psi$ ranges over all possible $M$.    See Theorem~\ref{thm-model-data}.   

Concerning the inverse problem,  the discrete invariant $A_\psi$ can be computed directly from the arrival time data $\sigma$, where $(\sigma,\alpha)$ is the data corresponding to some $(\tau,R)\in X_\psi$.   In practice the travel time vector $\tau$ and the matrix $A_\psi$ are simultaneously extracted from $\sigma$---see Algorithm~\ref{alg-inverse}.    The matrix $A_\psi$ then serves as a pointer, indicating which formulas $p_{\psi,n}(R)$ are associated with which amplitudes $\alpha_n$.  This allows one to pick out the amplitudes of primary reflections, which have the simplest formulas, and from which the reflection coefficients $R_n$ may be rapidly computed.  Crucially, it also allows one to  cross-check the reflection coefficients using the amplitudes of multiple reflections (a hugely redundant source) making it possible to correct amplitudes that may have been distorted.  The idea is implemented in Algorithm~\ref{alg-distorted}.  An important feature of the inverse algorithm (Algorithm~\ref{alg-inverse}), which depends on the irregular structure of generic models, is that it sequentially solves for $\tau$ from $\sigma$ and then for $R$ from $\alpha$.  (With one extra piece of data it is possible to solve the inverse problem for non-generic data, but the algorithm does not decouple as in the generic case and is therefore much slower.)   

This is not to say that the inverse problem can be solved in a practical way without limitation.   On the contrary, there is a simple geometric way to interpret the formula (\ref{arrival-times}) as a projection of lattice points that shows that for any precision $\varepsilon$, there are orientations of the vector $\tau$ (and hence associated $\psi$) for which successive entries in the arrival time data $\sigma$ will be separated by less than $\varepsilon$, rendering practical computation impossible.  See Section~\ref{sec-geometric}.

The key results in the present paper are: the explicit form of the amplitude polynomials $p_{\psi,n}$ (Theorem~\ref{thm-amplitude}),  the decoupled inverse algorithm (Algorithms~\ref{alg-inverse} and \ref{alg-distorted}), and the characterization of the mapping from models to data as locally linear/polynomial and globally injective (Theorems~\ref{thm-model-data} and \ref{thm-data-model}).

\subsection{Related literature\label{sec-literature}}

Newton's paper \cite{Ne:1980} surveys literature on the reflection problem up to 1980.   Apart from the papers cited already in previous sections, notable results since 1980 include those of Santosa-Schwetlick \cite{SaSc:1982} and Santosa-Symes \cite{SaSy:1988}.   The past two decades have seen considerable progress in the study of scaling limits for randomly layered media, and time reversal methods.  The book \cite{FoGaPaSo:2007}  by Fouque \emph{et al.} systematically covers the latter topics, which may be broadly characterized as stochastic.   By contrast, the present paper, while starting with the same layered medium as in \cite{FoGaPaSo:2007},  takes an entirely deterministic approach---and the analysis is entirely in the time domain.

\section{Solution to the forward problem\label{sec-forward}}

The main result of this section is Theorem~\ref{thm-amplitude}; the proof, which entails a comprehensive analysis of scattering sequences, is deferred to Section~\ref{sec-proof}.    The theorem requires a number of technical definitions that will also play a role in later sections.

\subsection{Key definitions: lattice set and amplitude polynomial\label{sec-key-definitions}}

The non-negative integer lattice $\integer^{M+1}_{\geq 0}$ plays a central role in the analysis of an $M$-layer model $(\tau,R)$; fix notation as follows.   A point $k\in\integer^{M+1}$ will be indexed starting with 0,
\[
k=(k_0,k_1,\ldots,k_M),
\]
consistent with the convention for models.  For lattice points $k,k^{\prime}\in\integer^{M+1}$, the notation $k\leq k^{\prime}$ refers to the entrywise comparison of vectors,
\[
k\leq k^\prime \mbox{ if and only if } k_n\leq k^{\prime}_n\mbox{ for each }0\leq n\leq M.
\]
(This will be referred to as the natural partial order on $\integer^{M+1}$.)   By the same token, the minimum of a pair of lattice points is to be interpreted entrywise,
\[
\min\{k,k^{\prime}\}=\bigl\{\min\{k_0,k^\prime_0\},\ldots,\min\{k_M,k^\prime_M\}\bigr\}.
\] 
The symbol $\mathbb{1}\in\integer^{M+1}$ indicates the vector of ones,
\[
\mathbb{1}=(1,1,\ldots,1),
\]
and for $0\leq n\leq M$, the symbol $k^n$ denotes the vector consisting of $n+1$ ones followed by $M-n$ zeros,
\begin{equation}\label{primary-defintion}
k^n_j=\left\{
\begin{array}{cl}
1&\mbox{ if }0\leq j\leq n\\
0&\mbox{ if }n+1\leq j\leq M
\end{array}.\right.
\end{equation}
So in particular $k^M=\mathbb{1}$ and $k^0=(1,0,\ldots,0)$; the lattice points $k^0,\ldots, k^{M}$ are called primary vectors.     
\begin{defn}[Lattice set]\label{defn-transit-count}
The set of $M$-layer transit count vectors is the subset 
\[
\mathfrak{L}_M\subset\integer^{M+1}_{\geq 0}
\]
consisting of all $k$ such that: (i) $k_0=1$; and (ii) for all $1\leq n\leq M$, if $k_n>0$ then $k_{n-1}>0$.   
Given an $M$-layer travel time sequence $\tau\in\real^{M+1}_{>0}$, define its lattice set as
\[
\mathfrak{L}^\tau_M=\left\{k\in\mathfrak{L}_M\,\bigl|\,\langle k,\tau\rangle\leq \langle\mathbb{1},\tau\rangle\right\}.
\]
\end{defn}
(The term ``transit count vector'' is a reference to scattering sequences; see Section~\ref{sec-proof}.)   Note that the lattice set $\mathfrak{L}^\tau_M$  of a travel time sequence is a finite set of lattice points, containing each of the primary vectors $k^0,\ldots,k^M$.  The notion of lattice set induces an equivalence relation on travel time sequences: declare $M$-layer travel time sequences $\tau$ and $\tau^\prime$ equivalent if $\LtM=\mathfrak{L}^{\tau^\prime}_M$.   The equivalence relation for $M=3$ is depicted in Figure~\ref{fig-equivalence} of Section~\ref{sec-hypothesis}.

The following (standard) multi-index notation helps to simplify some needed formulas.   Give a vector $x=(x_0,\ldots,x_M)$ and a lattice point $k\in\integer^{M+1}$,  $x^k$ denotes the monomial
\[
x^k=\prod_{n=0}^Mx_n^{k_n}.
\]
And given lattice points $b,k\in\integer^{M+1}$ with $0\leq b\leq k$, the multi-index binomial coefficient $\binom{k}{b}$ is defined by the formula
\[
\binom{k}{b}=\prod_{n=0}^M\binom{k_n}{b_n}=\prod_{n=0}^M\frac{k_n!}{b_n!(k_n-b_n)!}.
\]
\begin{defn}[Amplitude polynomial]\label{defn-amplitude-polynomial}
Given $M\geq 1$, let $x=(x_0,\ldots,x_M)$ denote a vector of variables; define auxiliary variables $y=(y_0,\ldots,y_M)$ by the formula 
\[
y_n=\sqrt{1-x_n^2}\quad\quad(0\leq n\leq M).
\]
For each $k\in\mathfrak{L}_M$, let $\tilde{k}$ denote its left shift,
$
\tilde{k}=(k_1,k_2,\ldots,k_M,0).
$
Set 
\[
u=\min\{\mathbb{1},\tilde{k}\}\quad \mbox{ and }\quad V(k)=\left\{b\in\integer^{M+1}\,\bigl|\,u\leq b\leq \min\{k,\tilde{k}\}\right\}.
\]
The $k$-amplitude polynomial $a(x,k)$ is defined as
\begin{equation}\label{amplitude-polynomial}
a(x,k)=\sum_{b\in V(k)}\binom{k}{b}\binom{\tilde{k}-u}{b-u}(-x)^{\tilde{k}-b}x^{k-b}y^{2b}.
\end{equation}
\end{defn}

The following propositions record properties and examples of amplitude polynomials that will be needed later.  
Observe that $a(x,k)$ has degree $2|k|-1$ 
and is homogenous with respect to the $2M+1$ variables $x_0,\ldots,x_M,y_0,\ldots,y_{M-1}$. (Since $b\leq\tilde{k}$ the variable $y_M$ does not occur.)   Homogeneity and odd degree imply that amplitude polynomials are odd, as follows.
\begin{prop}\label{prop-negative}
For every $k\in\mathfrak{L}_M$, \ $
a(-x,k)=-a(x,k).$
\end{prop}
\begin{pf} Since each $y_n$ has even degree in (\ref{amplitude-polynomial}), replacing $x$ with $-x$ is equivalent to replacing $(x,y)$ with $(-x,-y)$.   Homogeneity implies that this is equivalent to multiplication by $(-1)^{\deg a(x,k)}=-1$.\end{pf}\\

The following connection between amplitude polynomials of differing numbers of variables is immediate from the definition.  
\begin{prop}\label{prop-dimension}
Let $N\geq 1$.  Set $x=(x_0,\ldots,x_M)$ and $x^\prime=(x_0,\ldots,x_{M+N})$.  Let $k\in\mathfrak{L}_M$ and $k^\prime\in\mathfrak{L}_{M+N}$ be such that 
\[
k^\prime_n=\left\{
\begin{array}{cl}
k_n&\mbox{ if }0\leq n\leq M\\
0&\mbox{ if }M+1\leq n\leq M+N
\end{array}\right..
\]
Then $a(x,k)=a(x^\prime,k^\prime)$.   
\end{prop}
Thus an amplitude polynomial $a(x,k)$ depends only on the non-zero entries $k_n$ of $k$ and the corresponding variables $x_n$.   An important special case is that of primary vectors $k^n$, as defined above by (\ref{primary-defintion}).
\begin{prop}\label{prop-primary}
Let $1\leq n\leq M$, and let $k^n\in\integer^{M+1}$ denote the $n$th primary vector.  Then $V(k^n)=\bigl\{\widetilde{k^n}\bigr\}$ and 
\[
a(x,k^n)=x_n(1-x_0^2)(1-x_1^2)\cdots(1-x_{n-1}^2).
\]   
\end{prop}

A second example serves to illustrate how amplitude polynomials relate to one another when their respective transit count vectors differ in a single entry.   Fix notation as follows: for $0\leq n\leq M$, let $e^n$ denote the standard basis vector in $\integer^{M+1}$, whose entries are
\[
e^n_j=\left\{\begin{array}{cl}1&\mbox{ if }j= n\\
0&\mbox{ if }\quad j\neq n
\end{array}.\right.
\]
\begin{prop}\label{prop-redundancy}
Let $1\leq n\leq M-1$, and let $k,k^\prime\in\mathfrak{L}_M$ satisfy the conditions
\begin{equation}\label{redundancy}
\begin{split}
k_{n-1}=k_n=k_{n+1}=1\quad\mbox{ and }\quad k^\prime=k+e^n.
\end{split}
\end{equation}
Then 
\[
\frac{a(x,k^\prime)}{a(x,k)}=-2x_{n-1}x_n.
\]  
\end{prop}
\begin{pf}
Writing $u=\min\{\mathbb{1},\tilde{k}\}$ and $u^\prime=\min\{\mathbb{1},\tilde{k}^\prime\}$, observe that $u=u^\prime$ and furthermore that $V(k)=V(k^\prime)$.  Also, for each $b\in V(k)$,
\[
\begin{split}
&\binom{k^\prime}{b}=2\binom{k}{b},\quad\quad \binom{\tilde{k}^\prime-u}{b-u}=\binom{\tilde{k}-u}{b-u},\\[10pt]
 &(-x)^{\tilde{k}^\prime-b}=(-x_{n-1})(-x)^{\tilde{k}-b},\quad\mbox{ and }\quad x^{k^\prime-b}=x_nx^{k-b}.
 \end{split}
\]
In light of (\ref{amplitude-polynomial}) the conclusion of the proposition follows immediately.
\end{pf}

\subsection{Formula for the data in terms of the model\label{sec-formula}}

The impulse response may be expressed in terms of the underlying model using the amplitude polynomials of Definition~\ref{defn-amplitude-polynomial}, as follows. 
\begin{thm}\label{thm-amplitude}
Let $(\tau,R)$ be an arbitrary $M$-layer model.   Then
\begin{equation}\label{amplitude}
\gtr(t)=\sum_{k\in\mathfrak{L}_M}a(R,k)\delta(t-\langle k,\tau\rangle).
\end{equation}
\end{thm}
The proof of the theorem is postponed to Section~\ref{sec-proof}.  The simplicity of this result is remarkable---yet it is apparently new.   Restricting the sum (\ref{amplitude}) to the lattice set for $\tau$ adapts the result to the finite time interval $0\leq t\leq |\tau|$, as follows. 
\begin{cor}\label{cor-amplitude}
Let $(\tau,R)$ be an arbitrary $M$-layer model.   Then
\begin{equation}\label{data}
\chi_{[0,|\tau|]}\gtr(t)=\sum_{k\in\mathfrak{L}^\tau_M}a(R,k)\delta(t-\langle k,\tau\rangle).
\end{equation}
\end{cor}
Thus the data $(\sigma,\alpha)$ corresponding to a given model $(\tau,R)$ is obtained by putting the right-hand side of (\ref{data}) in normal form, thereby solving the forward problem.   It will be useful to formalize this procedure by introducing an explicit mapping, in order to distinguish some fundamentally different cases. 

\begin{defn}[Enumeration function]\label{defn-enumeration-function}
Given an $M$-layer model $(\tau,R)$, set $S=\supp\chi_{[0,|\tau|]}\gtr(t)$ and $d=\# S$.  The enumeration function of $(\tau,R)$, denoted $\psi=\Psi(\tau,R)$, is the mapping
\[
\psi:\LtM\rightarrow\{0,1,\ldots,d\}
\]
defined as follows.  If $\langle k,\tau\rangle\not\in S$, then set $\psi(k)=0$;   otherwise set 
\[
\psi(k)=1+\#\left\{ \sigma\in S\,\bigl|\,\sigma<\langle k,\tau\rangle\right\}.
\]
\end{defn}
The enumeration function $\psi=\Psi(\tau,R)$ need not be injective; in general it induces a partial ordering $\preceq$ on the lattice set $\LtM$ according to the prescription 
\begin{equation}\label{partial-ordering}
\forall k,k^\prime\in\LtM,\quad k\preceq k^\prime \Longleftrightarrow \psi(k)\leq\psi(k^\prime).
\end{equation}
The data for a given model may be expressed in terms of the model with the aid of its enumeration function as follows.  
\begin{cor}\label{cor-enumeration}
Given an $M$-layer model $(\tau,R)$, write 
\[
\chi_{[0,|\tau|]}\gtr(t)=\sum_{n=1}^d\alpha_n\delta(t-\sigma_n)
\]
in normal form.  Then, for $1\leq n\leq d$, 
\[
\sigma_n=\langle k,\tau\rangle\quad\mbox{ for any }k\in\psi^{-1}(n),
\]
and
\[
\alpha_n=\sum_{k\in\psi^{-1}(n)}a(R,k).
\]
\end{cor}
 
This solves the forward problem completely.   

\subsection{A geometric interpretation of arrival times\label{sec-geometric}}

\begin{figure}[h]
\begin{center}
\fbox{\includegraphics[width=.65\textwidth]{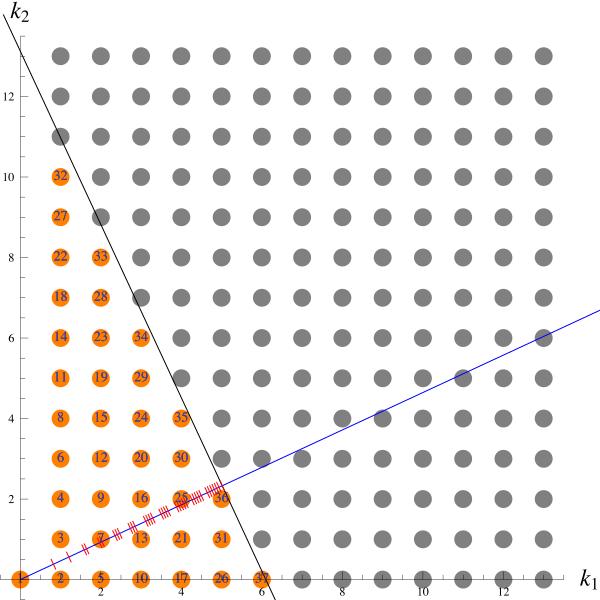}}
\caption{The lattice set $\mathfrak{L}^\tau_3$ (orange lattice points) along with $\mathfrak{L}_3$ (gray and orange lattice points), in the case $\tau=(1, 0.327971, 0.152455, 1.51957)$.  Vectors in $\mathfrak{L}^\tau_3$ have the form $k=(1,k_1,k_2,0)$, except $(1,1,1,1)$, which is omitted.  Points $k\in\mathfrak{L}_3^\tau$ are numbered according to $\psi(k)$.}
\label{figure-array}
\end{center}
\end{figure}

There is a simple geometric interpretation of the arrival times $\sigma_n$ in the typical case where $\psi$ is non-zero.  Up to rescaling, arrival times are essentially the orthogonal projection of the positive octant of the $M+1$-dimensional integer lattice onto a line having direction vector $\tau$.   More precisely, they are a rescaling of the projection of the lattice set $\LtM$ onto the line with direction vector $\tau$.  This can be sketched in low dimensions; as an illustration, consider the 3-layer case 
\[\tau=(1, 0.327971, 0.152455, 1.51957).\]
Except for the vector $k^3=(1,1,1,1)$, each member of $\mathfrak{L}^\tau_3$ lies in a two-dimensional plane in $\integer^4$.  
See Figure~\ref{figure-array}.  
For this particular choice of $\tau$, the numbers $\sigma_n$ corresponding to arrival time data are depicted in Figure~\ref{figure-array} as the orthogonal projection (red dashes) of the orange points onto the blue line.  The blue line is the orthogonal projection of $\tau$ onto the $(k_1,k_2)$-plane.  Later arrival times would correspond to projections of the gray points.  The numbering of the orange points indicates the values of the enumeration function $\psi$, which will be shown later to be a locally constant function of $\tau$.  

This view of arrival times as the orthogonal projection of lattice points onto a line explains both their pattern and density with increasing time.  It is closely related to the cut-and-project construction of quasicrystals due to Kramer and Neri \cite{KrNe:1984}.  In this connection see \cite{BaMo:2001} and also \cite{Fa:2012}, who traces lattice projections back to the almost periodic functions of Bohr \cite{Bo:1926}.  Note that for layered media the link to almost periodicity has a direct manifestation: the Fourier transform $\widehat{G^{(\tau,R)}}(\omega)$ of the impulse response is an almost periodic function in the sense of Bohr.  Returning to present considerations, as long as the direction of the line determined by $\tau$ is not rational, arrival times will be aperiodic and eventually arbitrarily dense.  On the other hand, the periodic structure of arrival times associated with, for example, a Bragg mirror, corresponds to a rational choice of $\tau$ that lines up with the lattice.

\section{Ill-posedness of the inverse problem\label{sec-ill-posed}}

\subsection{Examples of non-injectivity\label{sec-examples-non-injectivity}}

The mapping $(\tau,R)\mapsto(\sigma,\alpha)$ taking a model to its data is not injective, as the following example illustrates. 
\begin{example}\label{ex-non-injective}
Consider a 2-layer model of the form 
\[
(\tau,R)=\bigl((\tau_0,\tau_1,\tau_1),(R_0,\frac{1}{\sqrt{2}},R_0)\bigr),
\]
and the accompanying one-layer model 
\[
(\tau^\prime,R^\prime)=\bigl((\tau_0,\tau_1),(R_0,\frac{1}{\sqrt{2}})\bigr).
\]   
Evaluation of the formula in Corollary~\ref{cor-amplitude} shows that 
\[
\chi_{[0,|\tau|]}\gtr(t)=\chi_{[0,|\tau^\prime|]}G^{(\tau^\prime,R^\prime)}(t)=R_0\delta\bigl(t-\tau_0\bigr)+\textstyle\frac{1-R_0^2}{\sqrt{2}}\,\delta\bigl(t-(\tau_0+\tau_1)\bigr).
\]
Thus the two distinct models give rise to the same data. 
\end{example}
Let $\psi=\Psi(\tau,R)$, the enumeration function for the above $2$-layer model.   Notice that 
\[
\psi\bigl((1,2,0)\bigr)=\psi\bigl(k^2)=0.
\]
So $\psi$ both takes the value zero and fails to be injective.   Example~\ref{ex-non-injective} is part of a general pattern; similar examples exist in every higher dimension.   The following technical lemma captures the essential  reason.   
\begin{lem}\label{lem-non-injective}
Let $\tau$ be an $M+1$-layer travel time vector such that the set $S$, consisting of all $k\in\mathfrak{L}^\tau_{M+1}$ with the property that $k\neq k^{M+1}$ and 
$
\langle k,\tau\rangle=\langle k^{M+1},\tau\rangle,
$
has at least one element.   Let $\mathcal{R}$ denote the set of all points $(R_0,\ldots,R_M)\in(-1,1)^{M+1}$ for which there exists a non-zero value $R_{M+1}\in(-1,1)$ such that 
\begin{equation}\label{zero-coefficient}
a(R^\prime,k^{M+1})+\sum_{k\in S}a(R^\prime,k)=0,
\end{equation}
where $R^\prime=(R_0,\ldots,R_{M+1})$.  Then the Lebesgue measure of $\mathcal{R}$ in $\real^{M+1}$ is strictly positive.  
\end{lem}     
\begin{pf}   Note that if $k\in\mathfrak{L}^\tau_{M+1}$ is different from $k^{M+1}$, then $a(R^\prime,k)$ is independent of $R_{M+1}$.   Moreover, letting $n$ denote the maximal index such that $k_n\neq0$, the formula (\ref{amplitude}) shows that $a(R^\prime,k)$ is a multiple of $R_n$.  Therefore the set $V$ of points
\[
(R_0,\ldots,R_M)\in(-1/2,1/2)^{M+1}
\]
for which 
\[
0<\left|\sum_{k\in S}a(R^\prime,k)\right|\leq (3/4)^{M+1},
\]
has positive measure, since $0\not\equiv\sum_{k\in S}a(R^\prime,k)\rightarrow0$ as $\max_{0\leq n\leq M}|R_n|\rightarrow 0$.   
It follows from Proposition~\ref{prop-primary} that for each $(R_0,\ldots,R_M)\in V$, setting 
\[
R_{M+1}=-\sum_{k\in S}a(R^\prime,k)/\prod_{n=0}^M(1-R_n^2)
\]
yields the desired equation (\ref{zero-coefficient}).
\end{pf}

Many of the historical approaches to the 1-D seismic problem make a simplifying assumption that the travel time vector $\tau$ is constant.   But constant $\tau$ are easily seen to satisfy the hypothesis of Lemma~\ref{lem-non-injective}.  It follows from this that the inverse problem for such models is ill-posed.    
\begin{thm}\label{thm-ill-posed}
Let $\tau$ be a constant $M$-layer travel time vector, and let $\tau^\prime$ denotes its extension to an $M+1$-layer constant vector.  Write $\mathcal{R}\subset (-1,1)^{M+1}$ for the set of all points $R=(R_0,\ldots,R_M)$ for which there exists a non-zero $R_{M+1}\in(-1,1)$ such that the models $(\tau,R)$ and $(\tau^\prime,R^\prime)$ generate the same data, where $R^\prime=(R_0,\ldots,R_{M+1})$.    The set $\mathcal{R}$ has positive Lebesgue measure in $\real^{M+1}$.   
\end{thm}
\begin{pf}
Observe that $\tau^\prime$ satisfies the hypothesis of Lemma~\ref{lem-non-injective}; consider the reflectivities $R^\prime$ supplied by the lemma.  Note that $\mathfrak{L}^{\tau^\prime}_{M+1}\setminus\{k^{M+1}\}$ consists of the elements of $\LtM$, each extended by a 0 entry; therefore the supports of 
\[
\chi_{[0,|\tau|]}\gtr(t)\quad \mbox{ and }\quad \chi_{[0,|\tau^\prime|]}G^{(\tau^\prime,R^\prime)}(t)
\]
differ at most by $\langle k^{M+1},\tau^\prime\rangle$.  But the latter does not belong to the support of $\chi_{[0,|\tau^\prime|]}G^{(\tau^\prime,R^\prime)}(t)$, since its coefficient is given by the left side of equation~(\ref{zero-coefficient}) which, by Lemma~\ref{lem-non-injective}, is zero.  The remaining coefficents are identical between the two impulse responses, so $(\tau,R)$ and $(\tau^\prime,R^\prime)$ generate the same data.   \end{pf}

\subsection{The hypothesis of Lemma~\ref{lem-non-injective}\label{sec-hypothesis}}

What about non-constant travel time vectors?   After all, a randomly chosen $M$-layer travel time vector will not be constant.  (In fact, with probability one, it's entries will be linearly independent over the rational numbers.)  

\begin{figure}[h]
\parbox{.75\textwidth}{
\fbox{
\includegraphics[width=.75\textwidth]{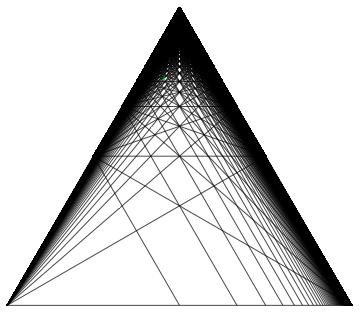}
}\\
\caption{The $3$-layer travel time vectors that satisfy Lemma~\ref{lem-non-injective}, represented by line segements within the triangle.  Each cell bounded by line segments represents the set of all vectors having a common lattice set.}\label{fig-equivalence}
}
\end{figure}

A starting point is to consider travel time vectors that satisfy the hypothesis of Lemma~\ref{lem-non-injective}, since this is the source of the ill-posedness expressed in Theorem~\ref{thm-ill-posed}.  The case of $3$-layer travel time vectors $\tau=(\tau_0,\tau_1,\tau_2,\tau_3)$ illustrates the general situation.  Observe that a transit count vector $k\in\LtM$ different from $k^3$ satisfies $\langle k,\tau\rangle=\langle k^3,\tau\rangle$ if and only if 
\begin{equation}\label{tau-hat}
(k_1-1)\hat{\tau}_1+(k_2-1)\hat{\tau}_2-\hat{\tau_3}=0,\quad\mbox{ where }\hat{\tau}=\textstyle\frac{1}{\tau_1+\tau_2+\tau_3}(\tau_1,\tau_2,\tau_3).
\end{equation}
Thus the normalized subvector $\hat\tau$ determines whether a given $3$-layer transit vector $\tau$ satisfies the hypothesis of Lemma~\ref{lem-non-injective}.  Of course $\hat\tau$ is a point in the standard two dimensional simplex $\Delta_2$, which is a triangle.   Each $k\in\mathfrak{L}_3$ determines a line segment in $\Delta_2$ defined by the equation (\ref{tau-hat}).    The set of all such lines is plotted in  Figure~\ref{fig-equivalence}. The vertices, clockwise from the top, are $\hat\tau=(0,0,1), (0,1,0)$ and $(1,0,0)$.  The union of lines represents the set of transit count vectors that satisfy the hypothsis of Lemma~\ref{lem-non-injective}, so the lemma fails to hold on the complement. The cells bounded by the lines represent the set of all travel time vectors that have a given lattice set, in the sense that $\mathfrak{L}_3^\tau$ is constant for $\hat\tau$ ranging over a single cell---with distinct cells corresponding to distinct lattice sets.   
The bottom left triangular cell represents the set of travel time vectors $\tau$ such that $\mathfrak{L}_2^\tau=\bigl\{k^0,k^1,k^2,k^3\bigr\}$, the minimum possible.  Moving to the right, each successive cell adds a vector of the form $k=(1,n,0,0)$, with $n$ increasing from $2$ to $\infty$.  Moving upward from the lower left, each successive cell adds a vector of the form $(1,1,n,0)$. 
\begin{figure}[h]
\parbox{.75\textwidth}{
\fbox{
\includegraphics[width=.75\textwidth]{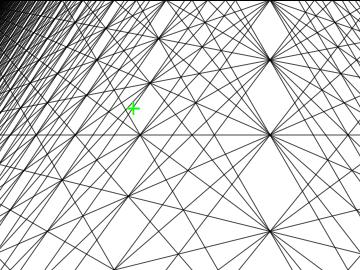}
}\\
\caption{A zoomed portion of the upper part of the triangle in Figure~\ref{fig-equivalence}. The green $+$ indicates the location of the travel time sequence depicted in Figure~\ref{figure-array}.}\label{fig-crop}
}
\end{figure}
A zoomed image from the upper part of the triangle is depicted in Figure~\ref{fig-crop}. 

Of course there is a corresponding construction in every higher dimension.  In general, for $M$-layer travel time vectors, the triangle $\Delta_2$ becomes a standard $(M-1)$-dimensional simplex $\Delta_{M-1}$, lines become hyperplanes, and polygonal cells become convex polytopes.   Lemma~\ref{lem-non-injective} fails on the interior of each of the convex polytopes.   

Thus in general Lemma~\ref{lem-non-injective} only holds on a set of $M$-layer travel time vectors of measure zero in $\real^{M+1}_{>0}$; this set happens to include constant travel time vectors.  The idea of the next section is to recover injectivity of the mapping from models to data by excluding a thin set of models which encompasses all those whose travel time vectors satisfy the hypothesis of Lemma~\ref{lem-non-injective}.

\section{The inverse problem for generic models\label{sec-generic}}

\subsection{Definition and characterization of generic models\label{sec-generic-definition}}

\begin{figure}[h]
\parbox{.75\textwidth}{
\fbox{
\includegraphics[width=.75\textwidth]{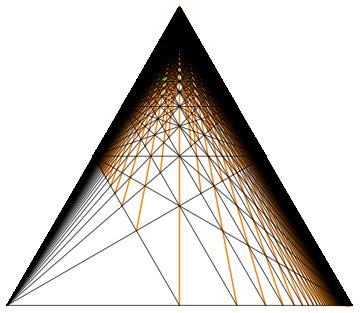}
}\\
\caption{The $3$-layer travel time vectors that fail to be generic, represented by line segements within the triangle.  The sets $U_\psi$ are represented by the open cells bounded by line segments.}\label{fig-generic}
}
\end{figure}

\begin{defn}[Generic]\label{defn-generic}
A model $(\tau,R)$ is termed generic if its enumeration function is injective and never zero.   
\end{defn}
Thus if $(\tau,R)$ is generic, then its enumeration function $\psi=\Psi(\tau,R)$ has an inverse,
\begin{equation}\label{psi-inverse}
\varphi=\psi^{-1}:\{1,\ldots,d\}\mapsto\LtM.
\end{equation}
Here is an alternate characterization that follows directly from Defintion~\ref{defn-enumeration-function}.  
\begin{prop}\label{prop-generic}
An $M$-layer model $(\tau,R)$ is generic if and only if the following conditions hold.
\begin{enumerate}
\item The mapping $k\mapsto\langle k,\tau\rangle$ is injective on $\LtM$.
\item For each $k\in\LtM$, $a(R,k)\neq 0$.   
\end{enumerate}
\end{prop}
It will be useful to refer to the reflectivity sequences that satisfy property (2) of Proposition~\ref{prop-generic}, and furthermore to have a notation for the set of travel time vectors associated to a common enumeration function. 
\begin{notation}\label{not-U-R}
Let $\psi=\Psi(\tau,R)$ for some generic $M$-layer model $(\tau,R)$.  Set 
\[
\mathcal{R}_\psi=\left\{R^\prime\in(-1,1)^{M+1}\;\bigl|\;a(R^\prime,k)\neq0\;\forall\;k\in\LtM\right\}.
\]
In addition, set 
\[
U_\psi=\left\{\tau^\prime\in\real^{M+1}_{>0}\;\bigr|\;\forall\;R^\prime\in\mathcal{R_\psi},\;\Psi(\tau^\prime,R^\prime)=\psi\right\}.
\]
\end{notation}
Observe that $\tau^\prime\in U_\psi$ only if both $\mathfrak{L}^{\tau^\prime}_M=\LtM$ and condition (1) of Proposition~\ref{prop-generic} holds.   More than this, 
\begin{prop}\label{prop-U}
If $\psi=\Psi(\tau,R)$, then $U_\psi\subset\real^{M+1}$ is a convex open set containing $\tau$, and $\mathcal{R}_\psi\subset(-1,1)^{M+1}$ is an open set containing $R$.
\end{prop}
As in Section~\ref{sec-hypothesis}, the case $M=3$ serves to illustrate the general situation.   Proposition~\ref{prop-generic} implies that the question of whether or not a given 
\[
\tau=(\tau_0,\ldots,\tau_3)\in\real^4_{>0}
\]
 is generic depends only on the normalized vector 
\[
\hat\tau=\textstyle\frac{1}{\tau_1+\tau_2+\tau_3}(\tau_1,\tau_2,\tau_3).
\]
In detail, $\tau$ fails to be generic if and only if there exist $k,k^\prime\in\mathfrak{L}_3$ such that 
\begin{subequations}\label{generic-cell}
\begin{align}
\langle(k_1,k_2,k_3),&\hat\tau\rangle\leq\langle\mathbb{1},\hat\tau\rangle,\quad\quad\langle(k^\prime_1,k^\prime_2,k^\prime_3),\hat\tau\rangle\leq\langle\mathbb{1},\hat\tau\rangle\label{in-the-same-cell}\\
\mbox{ and }\quad&\langle(k_1-k^\prime_1,k_2-k^\prime_2,k_3-k^\prime_3),\hat\tau\rangle=0.\label{same-ordering}
\end{align}
\end{subequations}
Note that the two inequalities (\ref{in-the-same-cell}) simply express that $k,k^\prime\in\mathfrak{L}^\tau_3$.   The particular case where $k\in\mathfrak{L}^\tau_3$ and $k^\prime=k^3$ reduces to equation (\ref{tau-hat}).   Therefore the set of $\hat\tau$ that satisfy (\ref{generic-cell}) for some $k,k^\prime\in\mathfrak{L}_3$ is a superset of the line segments plotted in Figure~\ref{fig-equivalence}.    The full set of solutions is plotted in Figure~\ref{fig-generic}, with the additional segments coloured orange.   The sets $U_\psi$ are the travel time vectors $\tau$ that correspond to the interiors of the cells in this figure.  
\begin{figure}[h]
\parbox{.75\textwidth}{
\fbox{
\includegraphics[width=.75\textwidth]{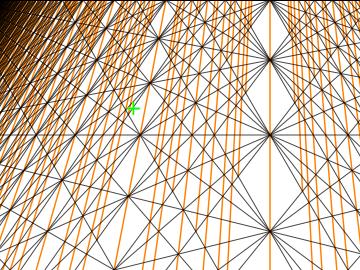}
}\\
\caption{A zoomed portion of the upper part of the triangle in Figure~\ref{fig-generic}. Again, the green $+$ indicates the travel time sequence of Figure~\ref{figure-array}.}\label{fig-generic-crop}
}
\end{figure}
A zoomed image from the upper part of the triangle is depicted in Figure~\ref{fig-generic-crop}. 

A similar picture exists in every higher dimension.   Thus in general, $U_\psi$ is the interior of the postive cone over a convex polytope crossed with $\real_{>0}$.  The important point is that $U_\psi$ is a convex open neighbourhood of $\tau$, and such neighbourhoods partition the set of all travel time vectors belonging to generic models.   

As for reflectivities, given a fixed $\psi=\Psi(\tau,R)$, Proposition~\ref{prop-generic} asserts that the vectors $R^\prime\in(-1,1)^{M+1}$ excluded from $\mathcal{R}_\psi$ are precisely those satisfying a polynomial equation of the form 
\[
a(R^\prime,k)=0\quad\mbox{ for some }\quad k\in\LtM.
\]
Thus $\mathcal{R}_\psi$ is the complement of a finite union of algebraic hypersurfaces in $(-1,1)^{M+1}$.  In particular, $\mathcal{R}_\psi$ is open and has full measure.

The enumeration function for a generic model may be represented as a matrix, as follows. 
\begin{defn}[Enumeration matrix]\label{defn-enumeration-matrix}
Let $(\tau,R)$ be a generic model, so that $\psi=\Psi(\tau,R)$ is a bijection of the form
\[
\psi:\LtM\rightarrow\{1,2,\ldots,d\}.
\]
The enumeration matrix representing $\psi$ is defined to be 
\[
A_\psi=\Bigl(\psi^{-1}(1)^T\;\;\psi^{-1}(2)^T\;\cdots\;\psi^{-1}(d)^T\Bigr). 
\]
Thus $A_\psi$ is the $(M+1)\times d$ integer matrix whose $n$th column is the transpose of the vector $\psi^{-1}(n)\in\LtM$.   
\end{defn}
Note that since each of the primary vectors $k^0,\ldots,k^M$ belongs to $\LtM$, the matrix 
\begin{equation}\label{K}
K_M=\Bigl((k^0)^T\;\;(k^1)^T\;\cdots\;(k^M)^T\Bigr)=
\left(\begin{array}{ccccc}
1&1&1&\cdots&1\\
0&1&1&\cdots&1\\
0&0&1&\cdots&1\\
\vdots&\vdots&\ddots &\ddots&\vdots\\
0&0&\cdots&0&1
\end{array}
\right)
\end{equation}
is a submatrix of $A_\psi$.   Therefore $A_\psi$ has rank $M+1$.    For future reference, let $J_M$ denote the inverse of $K_M$,
\begin{equation}\label{J}
J_M=K_M^{-1}=
\left(\begin{array}{cccccc}
1&-1&0&0&\cdots&0\\
0&1&-1&0&\cdots&0\\
0&0&1&-1&\ddots&\vdots\\
0&0&0&1&\ddots&0\\
\vdots&\vdots&\vdots&\ddots &\ddots&-1\\
0&0&0&\cdots&0&1
\end{array}
\right).
\end{equation}

\subsection{The model-data correspondence}

The notation developed in the previous section sets the stage for a very simple description of the correspondence between a generic model and its data.   
\begin{thm}\label{thm-model-data}
Fix an enumeration function $\psi$, and let $(\tau,R)\in U_\psi\times\mathcal{R}_\psi$ have associated data  
\[
(\sigma,\alpha)=\bigl((\sigma_1,\ldots,\sigma_d),(\alpha_1,\ldots,\alpha_d)\bigr).
\] 
Then the data is given in term of the model by the formulas
\begin{equation}\label{model-data}
\sigma=\tau A_\psi\quad\mbox{ and }\quad \alpha_n=a\bigl(R,\psi^{-1}(n)\bigr)\quad(1\leq n\leq d).
\end{equation}
\end{thm}
The crucial point is that the formulas (\ref{model-data}) hold locally; the same formula is valid throughout the open set $U_\psi\times\mathcal{R}_\psi$.   And open sets of the form $U_\psi\times\mathcal{R}_\psi$ partition the set of all generic $M$-layer models.   The inverse mapping, as it applies to data for generic models, may be characterized explicitly as follows.   

\begin{thm}\label{thm-data-model}
Let $(\sigma,\alpha)$ be the data corresponding to an $M$-layer model 
\[
(\tau,R)\in U_\psi\times\mathcal{R}_\psi,
\]
for some enumeration function $\psi$, and let $\sigma_\psi$ denote the subvector
\[
\sigma_\psi=\bigl(\sigma_{\psi(k^0)},\sigma_{\psi(k^1)},\ldots,\sigma_{\psi(k^M)}\bigr)\in\real^{M+1}.
\]
Then the model is given in terms of the data by the formulas
\begin{gather}
\tau=\sigma_\psi J_M,\label{data-model-times}\\[10pt]
R_0=\alpha_1, \qquad R_n=\frac{\alpha_{\psi(k^n)}}{\prod_{j=0}^{n-1}(1-R_j^2)}\quad(1\leq n\leq M)\label{data-model-reflectivities},
\end{gather}
where $J_M$ is the matrix (\ref{J}), and, as usual, $k^n$ denotes the $n$th primary travel time vector.   
\end{thm}
Thus the restriction of inverse map to the image of $U_\psi\times\mathcal{R}_\psi$ is linear in the arrival time data, by (\ref{data-model-times}).    The recursive scheme (\ref{data-model-reflectivities}) captures the precise non-linearity of the reflectivities, as determined by the amplitude data.   An important consequence of the inverse formulas is injectivity of the forward map.     
\begin{cor}\label{cor-injective}
Restricted to generic models, the mapping from models to data is injective.  
\end{cor}
\begin{pf}
Local injectivity of the forward map is implicit in Theorem~\ref{thm-data-model}, which gives the pre-image of a point in the range of a given patch $U_\psi\times\mathcal{R}_\psi$.   Global injectivity of the mapping $(\tau,R)\mapsto(\sigma,\alpha)$ follows from the inverse algorithm in Section~\ref{sec-algorithms}, Algorithm~\ref{alg-inverse}, which computes $\tau$---and hence $\psi$---directly from $\sigma$.   Thus if two models have the same image, they must belong to the same patch and therefore be identical.  
\end{pf}\\

\begin{figure}[h]
\parbox{4in}{
\fbox{
\includegraphics[width=3.75in]{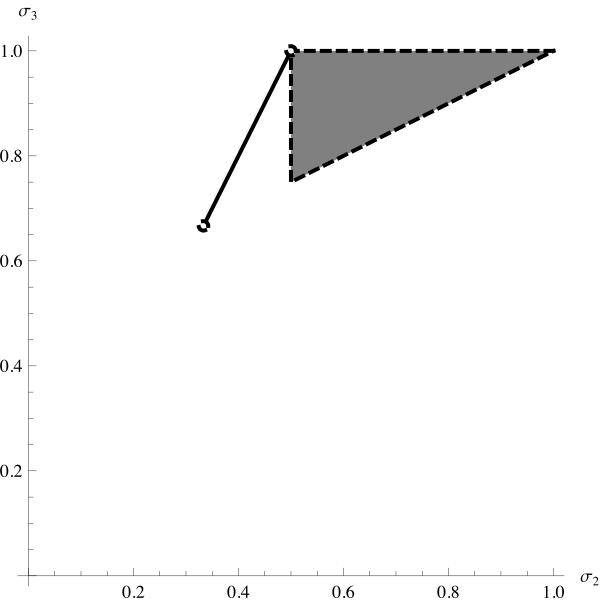}
}
\caption{The set of arrival time vectors $\sigma\in\real^4$ belonging to the range of the forward map, normalized and projected onto the $(\sigma_2,\sigma_3)$-plane.   The triangle is the image of 3-layer models, and the segment is the image of 2-layer models.}\label{fig-range}
}
\end{figure}

By Corolloary~\ref{cor-injective} the question of well-posedness of the inverse problem boils down to continuous dependence of the model on the data.   If one restricts the question to the image of a particular open set of generic models $U_\psi\times\mathcal{R}_\psi$, there is no problem.  The forward and inverse maps are diffeomorphisms.   Considered globally, however, the situation is much more delicate.   To take a concrete example, consider a point 
\[
\sigma=(\sigma_1,\sigma_2,\sigma_3,\sigma_4)\in\real^4
\]
as a candidate for travel time data.   Is $\sigma$ in the range of the forward mapping?   And if so, what is the structure of the set of nearby data points?    By Theorem~\ref{thm-data-model}, the set of arrival time vectors $\sigma\in\real^d$ belonging to the range of the forward map is a finite union of convex polytopes of various dimensions $\leq d$, each of which is open relative to its affine hull.    The case $d=4$ is illustrated in Figure~\ref{fig-range}, where the data is projected onto the $(\sigma_2,\sigma_3)$-plane, and normalized so that it lies inside the unit square.   
There are two separate pieces, a triangle, which corresponds to $3$-layer models whose data has 4 terms, and a segment, which corresponds to $2$-layer models whose data has 4 terms.   The intersection of the closures of these two pieces is a single point that is just like Example~\ref{ex-non-injective}. It is clear even in this low dimensional case that an arbitrary point $\sigma\in\real^4$ will not in general have a unique closest point in the closure of the range of the forward mapping.  Thus there is no clear-cut way to project a given candidate for a data point onto the range.   (See Section~\ref{sec-spurious} for a different approach.)   Thus the inverse problem is well-posed if one restricts to the range of the forward mapping; but the range itself has an irregular structure.

\section{Algorithms and examples\label{sec-algorithms}}

With an explicit formula for the truncated impulse response 
\[
\chi_{[0,|\tau|]}\gtr(t)=\sum_{n=1}^d\alpha_n\delta(t-\sigma_n)
\]
in terms of the underlying model $(\tau,R)$, it is a straighforward matter to compute the exact impulse response efficiently, provided the number of layers is not so large as to render the inherent polynomial evaluations prohibitively expensive.  The corresponding inverse algorithm, while perhaps less obvious, does not require the polynomial evaluations inherent in the forward algorithm.    Rather, it is constrained by search and sort procedures, details of which are given in Section~\ref{sec-inverse}.

\subsection{Forward algorithm\label{sec-forward}}
The following algorithm encodes the solution to the forward problem, using formula (\ref{aRk}) of Theorem~\ref{thm-amplitude}.
\begin{alg}[Forward]\label{alg-forward}\ \\[-20pt]
\begin{tabbing}
In\=put:\ \  $(\tau,R)=\left((\tau_0,\ldots,\tau_M),(R_0,\dots,R_M)\right)$\\[5pt]
\> Step 1: Allocate $\LtM=\left(k^{(1)},\ldots,k^{(d)}\right)$.\\[5pt]
\> Step 2: Evaluate $\tau\cdot\LtM=\left(\langle\tau,k^{(1)}\rangle,\ldots,\langle \tau,k^{(d)}\rangle\right)$.\\[5pt]
\> Step 3: Sort $\tau\cdot\LtM$ into increasing order, $\sigma=\left(\langle\tau,k^{(\pi(1))}\rangle,\ldots,\langle \tau,k^{(\pi(d))}\rangle\right)$.\\[5pt]
\> Step 4: Compute $\alpha_n=a(R,k^{(\pi(n))})$ for $n=1,\ldots,d$, using formula (\ref{aRk}). \\[5pt]
Output: $(\sigma,\alpha)=\left((\sigma_1,\ldots,\sigma_d),(\alpha_1,\ldots,\alpha_d)\right)$ 
\end{tabbing}
\end{alg}
The output of the algorithm is the arrival times and amplitudes for the normal form of $\gtr(t)$, restricted to $t\leq|\tau|$,
\[
\chi_{[0,|\tau|]}\gtr(t)=\sum_{n=1}^d\alpha_n\delta(t-\sigma_n).
\]

\subsubsection{An example with 17 reflectors\label{sec-17}}
\begin{figure}[h]
\parbox{5in}{
\fbox{
(a)\ \includegraphics[width=2.15in]{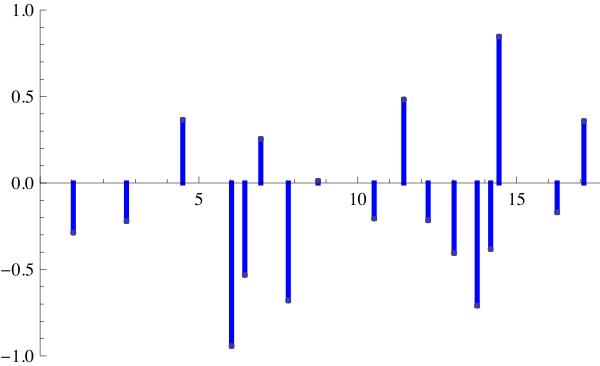}
(b)\ \includegraphics[width=2.15in]{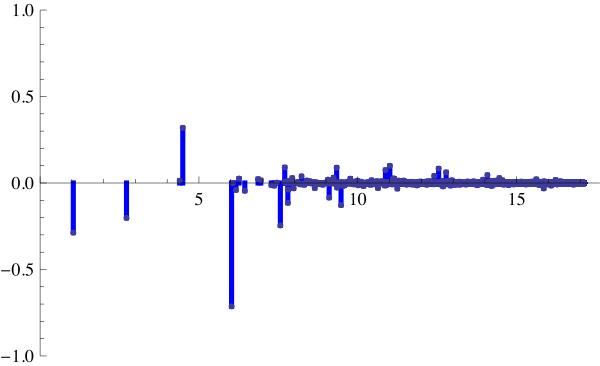}
}\\
\caption{A 16 layer model (i.e. with 17 reflectors).  (a) The true model.  (b) The corresponding impulse response computed by Algorithm~\ref{alg-forward}.}\label{fig-forward}
}
\end{figure}

Figure~\ref{fig-forward} depicts a 17-layer model (where $R_n$ is plotted against $\tau_0+\cdots+\tau_n$), and its impulse response computed by Algorithm~\ref{alg-forward} (in approximately 17.1 seconds).   This impulse response has 33714 terms, most of which are low amplitude late arrivals, invisible in Figure~\ref{fig-forward} (b).  Note the very small amplitude $0.012$ term at $8.75$ seconds of the model.   While the first four terms in the model have easily recognized corresponding primary reflections in the impulse response, this small amplitude is totally obscured.  There is no way to recognize its associated primary reflection by simply studying the plot of the impulse response, no matter what the scale.   

\begin{figure}[H]
\parbox{5in}{
\fbox{
(a)\ \includegraphics[width=2.15in]{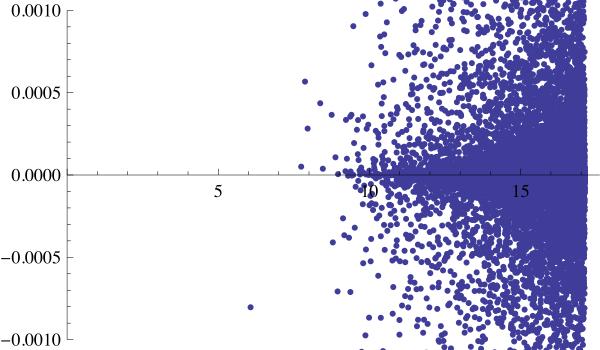}
(b)\ \includegraphics[width=2.15in]{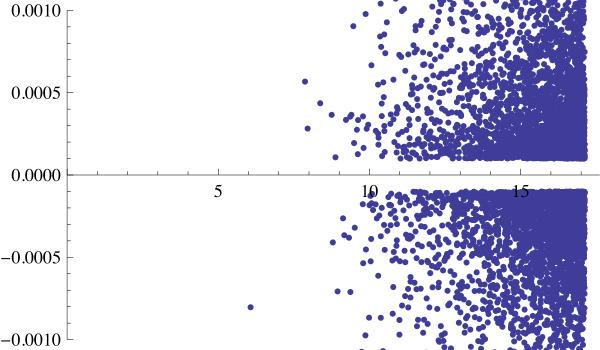}
}\\
\caption{A zoomed version of the impulse response.  (a) The true impulse response (plotted without stems), and (b) a decimated version. }\label{fig-zoom}
}
\end{figure}

Figure~\ref{fig-zoom} (a) shows a zoomed view of the impulse response (plotted without stems) revealing the explosion of later arrivals.   Figure~\ref{fig-zoom} (b) shows a 7003 term decimated version of the impulse response in which all terms having absoute amplitude $<10^{-4}$ have been removed.  

\subsection{Inverse algorithm\label{sec-inverse}}

For the inverse algorithm, a further subdivision of $\LtM$ is needed.   Given $\tau^{(n)}=(\tau_0,\ldots,\tau_n)$ and $s>0$, write
\[
\Ltns=\left\{k\in\mathfrak{L}_n\,\left|\,k_n\geq1\mbox{ and }\langle \tau^{(n)},k\rangle\leq s\right.\right\}.
\]

\begin{alg}[Inverse]\label{alg-inverse}\ \\[-20pt]

\begin{tabbing}
In\=put:\ \ $(\sigma,\alpha)=\left((\sigma_1,\ldots,\sigma_d),(\alpha_1,\ldots,\alpha_d)\right)$ \\[5pt]
{Stage I: Arrival time inversion}\\[5pt]
\> Step 1:\ \ Set $(\tau_0,\tau_1)=(\sigma_1,\sigma_2-\sigma_1)$, $S=\{\sigma_j\,|\,3\leq j\leq d\}$ and $n=1$.\\[5pt]
\> St\=ep 2a:\ \ Given $\tau^{(n)}=(\tau_0,\ldots,\tau_n)$, allocate $\Ltn$.\\[5pt]
\>\>Step 2b:\ \ Set $S=S\setminus\left\{\left. \langle\tau^{(n)},k\rangle\,\right|\,k\in\Ltn\right\}$; if $S=\emptyset$, go to Step~3.\\[5pt]
\>\>Step 2c:\ \ Set $\tau_{n+1}=\min S-|\tau^{(n)}|$ and $n=n+1$; return to Step~2a. \\[5pt]
\> Step 3: Set $M=n$, $\tau=\tau^{(n)}$.\\[5pt]
{Stage II: Amplitude inversion}\\[5pt]
\> Step 4:  For each $0\leq N\leq M$, determine $\rho(N)$ such that $\sigma_{\rho(N)}=\langle\tau,k^N\rangle$.\\[5pt]
\> Step 5:  Set $R_0=\alpha_{\rho(0)}$.\\[5pt]
\> Step 6: For each $1\leq N\leq M$, set $R_{N}=\frac{\alpha_{\rho(N)}\cdot R_{N-1}}{\alpha_{\rho(N-1)}\cdot (1-R_{N-1}^2)}$.\\[5pt]
Output: $(\tau,R)=\left((\tau_0,\ldots,\tau_M),(R_0,\dots,R_M)\right)$
\end{tabbing}
\end{alg}

Note that Step~6 uses the result detailed in Proposition~\ref{prop-primary}.  The above inverse algorithm has a number of remarkable properties, as follows.
\begin{itemize}
\item The algorithm is exact.  
\item The arrival time inversion is independent of amplitude data.  
\item If the input arrival times are collectively shifted by $\kappa$, giving modified input 
\[ (\sigma+\kappa,\alpha)=\left((\sigma_1+\kappa,\ldots,\sigma_d+\kappa),(\alpha_1,\ldots,\alpha_d)\right),
\]
then the resulting output is $(\tilde{\tau},R)=\left((\tau_0+\kappa,\tau_1,\ldots,\tau_M),(R_0,\dots,R_M)\right)$.
\item Given any subsequence of the original input data that contains all the primary reflections, 
\[
(\tilde{\sigma},\tilde{\alpha})=\left((\sigma_{m_1},\ldots,\sigma_{m_n}),(\alpha_{m_1},\ldots,\alpha_{m_n})\right),
\]
the resulting output is $(\tau,R)$.   In other words, the algorithm recovers the model exactly given only partial data. 
\item The algorithm is fast, and can be accelerated by deleting low amplitude data, provided no primary reflections are deleted.     
\end{itemize} 

The inverse algorithm recovers the 16-layer model depicted in Figure~\ref{fig-forward} from its impulse response exactly; the computation took 127 seconds in a less-than-optimal implementation in Mathematica.   Remarkably, Algorithm~\ref{alg-inverse} also correctly recovers the model from its decimated impulse response depicted in Figure~\ref{fig-zoom}; in this case the computation took just 28.2 seconds.   In both cases the recovered model is identical with the plot in Figure~\ref{fig-forward}(a) (including the small amplitude term at 8.75 seconds!).   

\begin{figure}[h]
\parbox{5in}{
\fbox{
(a)\ \includegraphics[width=2.15in]{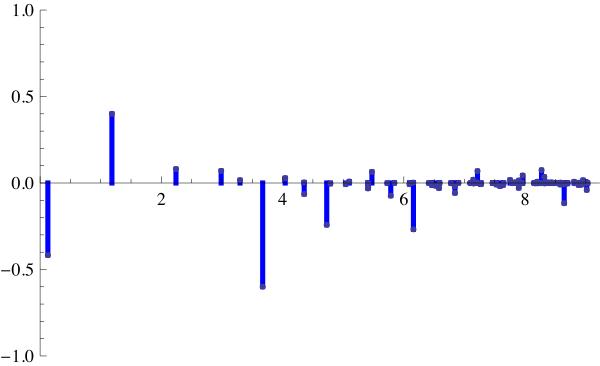}
(b)\ \includegraphics[width=2.15in]{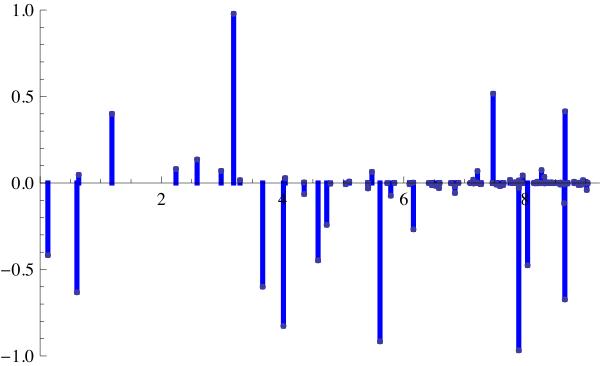}
}\\
\caption{(a) The true impulse response for a 7-layer model (115 data points), and (b) with 12 spurious arrivals added.}\label{fig-seven}
}
\end{figure}

\begin{figure}[h]
\parbox{5in}{
\fbox{
(a)\ \includegraphics[width=2.15in]{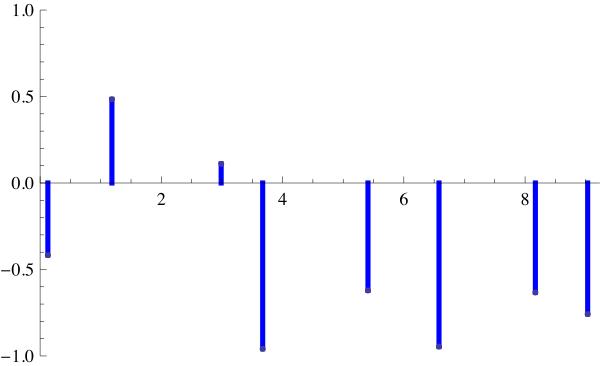}
(b)\ \includegraphics[width=2.15in]{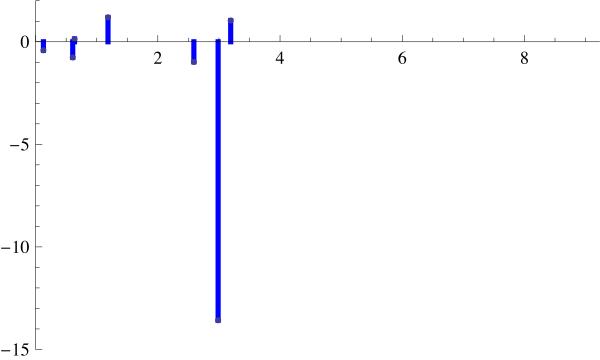}
}\\
\caption{(a) The true 7-layer model corresponding to Figure~\ref{fig-seven}.  (b) The partial model computed from by Algorithm~\ref{alg-inverse} with the corrupted data as input; the computation was terminated after 612 seconds.}\label{fig-failure}
}
\end{figure}

Figure~\ref{fig-seven} depicts the impulse response for a 7-layer model, with some randomly generated spurious arrivals added on top.  While Algorithm~\ref{alg-inverse} recovers the model exactly from the uncorrupted data, it falters when fed the data that includes spurious arrivals.   See Figure~\ref{fig-failure}.

\subsubsection{Modification to detect spurious arrivals\label{sec-spurious}}   The inverse algorithm can be modified to handle the presence of spurious arrivals.   One way to control for this is the following modification to Step~2a.
\begin{tabbing}
In\=\kill\\
\>St\=ep~2a(i):\ \ Given $\tau^{(n)}=(\tau_0,\ldots,\tau_n)$, allocate $\Ltn$.\\[5pt]
\>\>Step~2a(ii):\ \ If $S\cap\left\{\left. \langle\tau^{(n)},k\rangle\,\right|\,k\in\Ltn\right\}=\left\{\min S\right\}\neq S$, then\\ 
\>\>\ \ set $S=S\setminus\{\min S\}$, set $\tau_n=\min S-|\tau^{(n-1)}|$, and return to Step~2a(i).
\end{tabbing}
The effect of this modification is to delete data points that appear to correspond to primary reflections, but for which there are no corresponding multiple reflections in the later data points.  The modified Algorithm~\ref{alg-inverse}, with the corrupted version of the above impulse response as input, recovers the model in Figure~\ref{fig-failure}(a) exactly. The computation took just $0.0569$ seconds.

\subsubsection{Modification to correct for distorted amplitudes\label{sec-distorted}}

Further, one can exploit redundancy of multiples to control for inaccurate amplitude data.  The idea is to use the formula (\ref{amplitude-polynomial}) to solve for products and ratios of reflection coefficients that are given by distinct data points.   There are many ways to do this.   As a simple illustration of the idea, the following scheme uses Proposition~\ref{prop-redundancy}.   Suppose that the above inverse algorithm has input $(\sigma,\alpha)$, where the amplitude $\alpha$ is distorted, and let $(\tau,R^\prime)$ denote the resulting output---where the $^\prime$ indicates that the reflectivity is distorted.  Assuming that $\tau$ is accurate, the following third stage is designed to follow Algorithm~\ref{alg-inverse}.  

\begin{alg}[Reflectivity correction]\label{alg-distorted}\ \\[-20pt]

\begin{tabbing}
In\=put:\ \ \=$(\tau,R^\prime)=\left((\tau_0,\ldots,\tau_M),(R^\prime_0,\dots,R^\prime_M)\right)$, and $(\sigma,\alpha)$, the respective\\[5pt]
\> output from, and input to, Algorithm~\ref{alg-inverse} \\[5pt]
{Stage III: Reflectivity correction}\\[5pt]
\> Step 7:  For each $1\leq n\leq M-3$, allocate the set $E_n$ of all pairs $(k,k^\prime)$\\[5pt] 
\>\> from $\LtM$ satisfying condition (\ref{redundancy}) in Proposition~\ref{prop-redundancy}.\\[5pt]
\> Step 8:  For each $1\leq n\leq M-3$, compute  the set $C_n$ of all ratios $-\frac{\alpha_{j^\prime}}{2\alpha_j}$\\[5pt]
\>\>such that $\sigma_j=\langle k,\tau\rangle$ and $\sigma_{j^\prime}=\langle k^\prime,\tau\rangle$, where $(k,k^\prime)\in E_n$.\\[5pt]
\> Step 9: For each $1\leq n\leq M-3$, set $c_n$ to be the (mean of the) most\\[5pt]
\>\>common value(s) in $C_n$.\\[5pt]
\> Step 10: Set $R^{\prime\prime}_0=R^\prime_0$.  For $n$ from $1$ to $M-4$, set $R^{\prime\prime}_n=c_n/R^{\prime\prime}_{n-1}$.\\[5pt]
\> Step 11:  For $n$ from $M-3$ to $M$, let $j_n$ be such that $\sigma_{j_n}=\langle k^n,\tau\rangle$ (with \\[5pt]
\>\>  $k^n$ as in Proposition~\ref{prop-primary}), and set $R^{\prime\prime}_n=\alpha_{j_n}/\prod_{i=0}^{n-1}\bigl(1-(R^{\prime\prime}_i)^2\bigr)$.\\[5pt]
Output: $R^{\prime\prime}=(R^{\prime\prime}_0,\dots,R^{\prime\prime}_M)$
\end{tabbing}
\end{alg}

\begin{figure}[h]
\parbox{5in}{
\fbox{
(a)\ \includegraphics[width=2.15in]{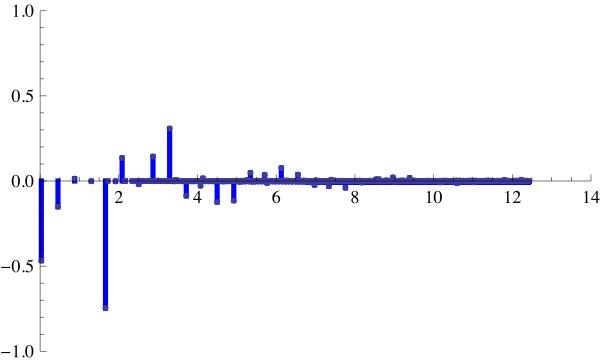}
(b)\ \includegraphics[width=2.15in]{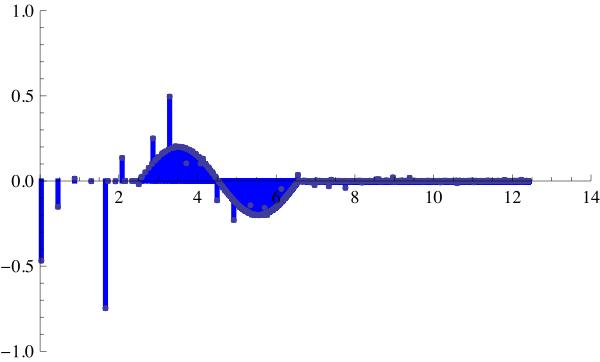}
}\\
\fbox{
(c)\ \includegraphics[width=2.15in]{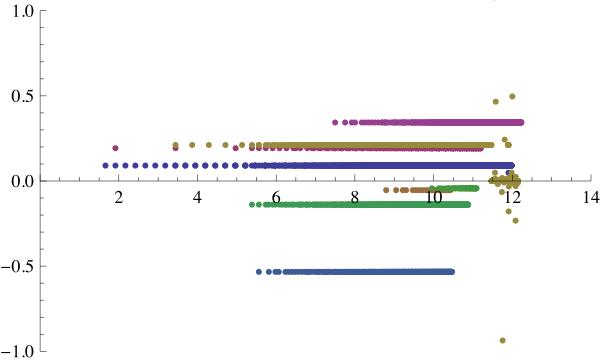}
(d)\ \includegraphics[width=2.15in]{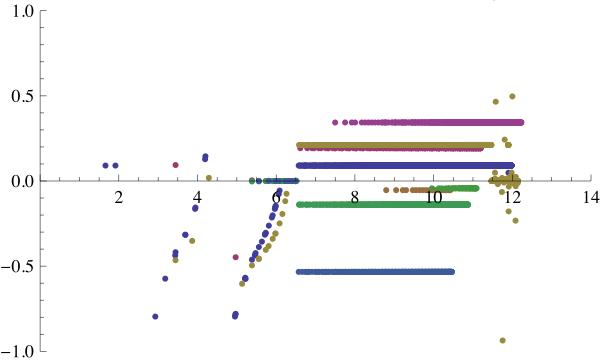}
}
\caption{An 11 layer model with distorted amplitudes.  (a) The true impulse response (27108 data points). (b) A $0.2$ amplitude  sine wave distortion added to the true impulse response between $t=2.51s$ and $t=6.56s$.  The sets $C_1,\ldots,C_9$ are plotted for the true impulse response (c) and the distorted impulse response (d).  The range of distortion is clearly evident.}\label{fig-distortion}
}
\end{figure}

The above reflectivity correction requires that the input value $R^\prime_0$ be accurate---meaning that $R^\prime_0=R_0$, the true reflectivity---and that the later amplitudes $\alpha_{j_n}$ invoked in Step~11 also be accurate.   As long as these assumptions are satisfied Algorithm~\ref{alg-distorted} can recover the true reflectivities.   Figure~\ref{fig-distortion} shows an example.  The short-term sine wave distortion of the amplitude is meant to mimic the phenomenon of ground roll endemic in land-based seismic acquisition.   

\begin{figure}[h]
\parbox{5in}{
\fbox{
(a)\ \includegraphics[width=2.15in]{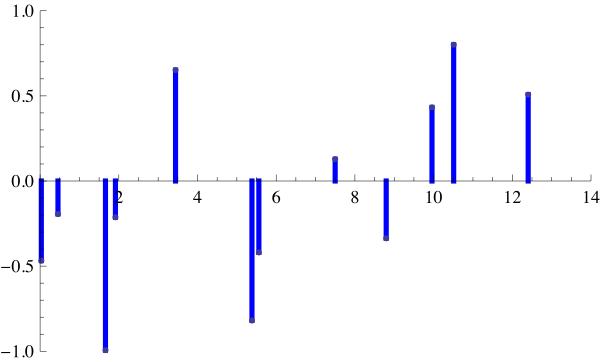}
(b)\ \,\includegraphics[width=2.15in]{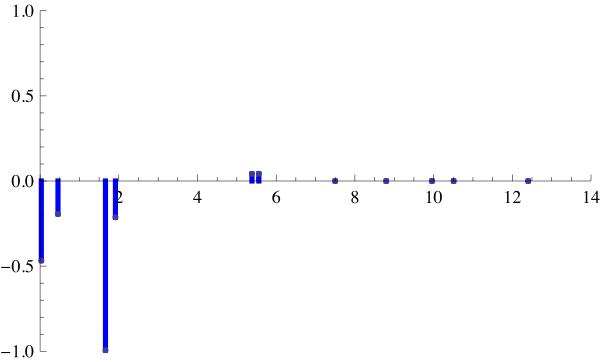}
}\\
\fbox{
(c)\ \includegraphics[width=2.15in]{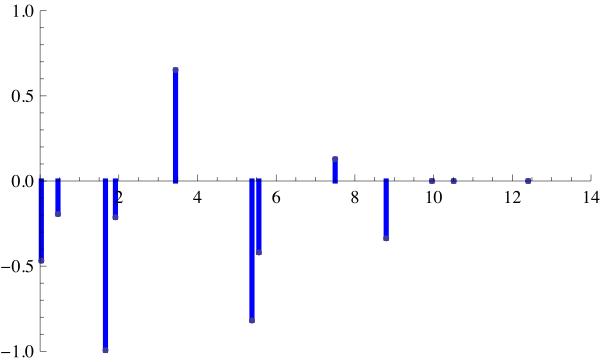}
(d)\ \includegraphics[width=2.15in]{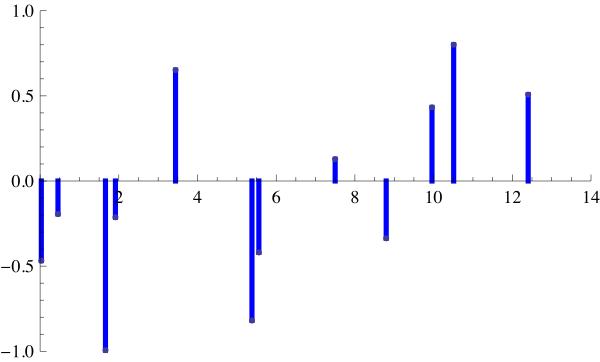}
}
 \caption{Amplitude correction applied to the impulse response for the 11 layer model of Figure~\ref{fig-distortion}.  The true model (a) and the model computed by Algorithm~\ref{alg-inverse} (b).  The correction after Step~10 of Algorithm~\ref{alg-distorted} (c) and after Step~11 (d).  The true model is recovered exactly. }\label{fig-distortion2}
}
\end{figure}

The results of Algorithm~\ref{alg-inverse} followed by Algorithm~\ref{alg-distorted}  are depicted in Figure~\ref{fig-distortion2}.   Despite the substantial sine wave distortion, the exact reflection coefficients are extracted using the set $C_n$.   Without Algorithm~\ref{alg-distorted}, only the first four reflection coefficients would have been correctly estimated.  

\section{Proof of Theorem~\ref{thm-amplitude}\label{sec-proof}}

A standard idea is to view $\gtr(t)$ as the sum total of all possible sequences of successive reflections and transmissions of the initial pulse that eventually return to $z_{-1}$.  The sequence of transmission coefficients $T=(T_0,\ldots,T_M)$ corresponding to a given sequence of reflection coefficients $R=(R_0,\ldots,R_M)$ is given by the formulas
\[
T_n=\sqrt{1-R_n^2}\quad\quad(0\leq n\leq M).
\]
(See \cite[Chapter~3]{FoGaPaSo:2007} for details.)  The idea is worth illustrating, since it underpins the arguments below. For example, after $\tau_0/2$ seconds, the initial downward traveling pulse $\delta(z-z_{-1}-c_0t)$ satisfying equations (\ref{wave1},\ref{wave2},\ref{initial}) reaches the interface at $z_0$.   Part of it is transmitted into the first layer as $T_0\delta\bigl(z-z_0-c_1(t-\frac{\tau_0}{2})\bigr)$, which, after another $\tau_1/2$ seconds, reaches the interface $z_1$.   Part of this pulse is reflected back into first layer as $R_1T_0\delta\bigl(z-z_1+c_1(t-\frac{\tau_0+\tau_1}{2})\bigr)$.   Traveling back up to $z_0$, the latter reaches $z_0$ after a further $\tau_1/2$ seconds, and is partly transmitted back to $z_{-1}$ as 
\begin{equation}\label{samplewave}
R_1T_0^2\delta\bigl(z-z_0+c_0(t-\textstyle\frac{\tau_0+2\tau_1}{2})\bigr),
\end{equation}
arriving at $z_{-1}$ at time $\sigma=\tau_0+\tau_1$.   Thus the part of the initial pulse that traverses the sequence of depths $\mpp=(z_{-1},z_0,z_1,z_0,z_{-1})$ returns to $z_{-1}$ with an amplitude $\alpha=R_1T_0^2$ at arrival time $\sigma=\tau_0+\tau_1$, thereby contributing a term of the form $\alpha\delta(t-\sigma)$ to $\gtr(t)$.   The sequence $\mpp$ is called a scattering sequence, and the amplitude $\alpha$ is called the weight of $\mpp$.   The impulse response itself is a delta train of the form
\begin{equation}\label{normalform}
G^{(\tau,R)}(t)=\sum_{n=1}^\infty\alpha_n\delta(t-\sigma_n),
\end{equation}
composed of the cumulative contributions of all possible scattering sequences returning to $z_{-1}$, with their associated weights and arrival times.  In the above example, $\mpp$ is the only scattering sequence having arrival time $\sigma$.  But in general different scattering sequences may arrive simultaneously, so each amplitude $\alpha_n$ occurring in (\ref{normalform}) is the sum of the weights of all scattering sequences arriving at time $\sigma_n$.   

While the interpretation of the impulse response in terms of scattering sequences as outlined above is well established, it has not been used as a means to express $\gtr(t)$ directly in terms of the model $(\tau,R)$.  This is perhaps because a full analysis of all possible scattering sequences is viewed as being enormously complicated, with any resulting representation being hopelessly cumbersome and of little practical use.   Such a view seems unjustified.   It is shown in the present section, that, with the right perspective, a direct analysis of scattering sequences leads to the surprisingly simple, explicit formula for $\gtr(t)$ stated in Theorem~\ref{thm-amplitude}.

\subsection{Scattering sequences and the impulse response}

The first step is to establish some terminology and notation, as follows.  A scattering sequence that starts and ends at $z_{-1}$ is represented by a path in the graph
\begin{equation}\label{simplegraph}
\mbox{
\includegraphics[width=4in]{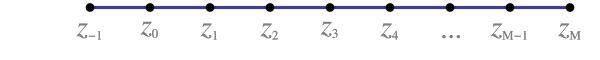}
}
\end{equation}
Formally, given an integer $M\geq 1$, let $\sss_M$ denote the set of all sequences of the form
\[
\mpp=(\mpp_0,\mpp_1,\ldots,\mpp_L)
\]
such that $L\geq 2$ and: 
\begin{subequations}
\begin{align}
\mpp_0=\mpp_L=z_{-1}\mbox{ and }\mpp_n\in&\left\{z_1,\ldots,z_M\right\}\mbox{ if }1\leq n\leq L-1;\label{pathcondition1}\\
\begin{split}
\forall n \mbox{ with } 0\leq n\leq L-1,&\quad\exists j \mbox{ with }-1\leq j\leq M-1\mbox{ such that }\\
&\{\mpp_n,\mpp_{n+1}\}=\{z_j,z_{j+1}\}.\label{pathcondition2}
\end{split}
\end{align}
\end{subequations}
The elements of $\sss_M$ will be referred to as scattering sequences.  Condition (\ref{pathcondition1}) says a scattering sequence starts and ends at $z_{-1}$, and condition (\ref{pathcondition2}) (which refers to \emph{un}ordered pairs) says that adjacent terms in a scattering sequence are adjacent vertices in the graph (\ref{simplegraph}). 

For example, for $0\leq n\leq M$, the shortest scattering sequence that reaches $z_n$ is  
\begin{equation}\label{primary}
(z_{-1},z_0,z_1,\ldots,z_{n-1},z_n,z_{n-1},\ldots,z_1,z_0,z_{-1});
\end{equation} 
this is called a primary scattering sequence.  A scattering sequence reaching maximum depth $z_n$ that is not shortest possible is called a multiple scattering sequence.

\subsubsection{The weight of a scattering sequence}

The weight $w(\mpp)$ corresponding to a scattering sequence 
\[
\mpp=(\mpp_0,\mpp_1,\ldots,\mpp_L)
\]
in an $M$-layer model $(\tau,R)$ is defined as follows.   For each $n$ in the range $1\leq n\leq L-1$, and given that $\mpp_n=z_j$, define
 \begin{equation}\label{wn}
 w_n=\left\{\begin{array}{cc}
 R_j&\mbox{ if }\mpp_{n-1}=\mpp_{n+1}=z_{j-1}\\
 -R_j&\mbox{ if }\mpp_{n-1}=\mpp_{n+1}=z_{j+1}\\
 \sqrt{1-R_j^2}&\mbox{ otherwise }
 \end{array}\right..
 \end{equation}
 The three possibilities correspond respectively to: reflection inside the $j$th layer at $z_j$; reflection inside the $(j+1)$st layer at $z_j$; and transmission between the $j$th and $(j+1)$st layers.  Finally, set 
 \begin{equation}\nonumber
 w(\mpp)=\prod_{n=1}^{L-1}w_n.
 \end{equation}
Thus the part of an initial unit impulse that traverses $\mpp$ returns to $z_{-1}$ with amplitude $w(\mpp)$.

\subsection{Transit count and branch count vectors}

The purpose of this section is to define two maps, 
\[
\kappa,\beta:\sss_M\rightarrow\integer^{M+1}
\]
that associate integer vectors to a given scattering sequence.    

\begin{figure}[h]
\fbox{
\includegraphics[clip,trim=0in 1in 0in 0in, width=4.8in]{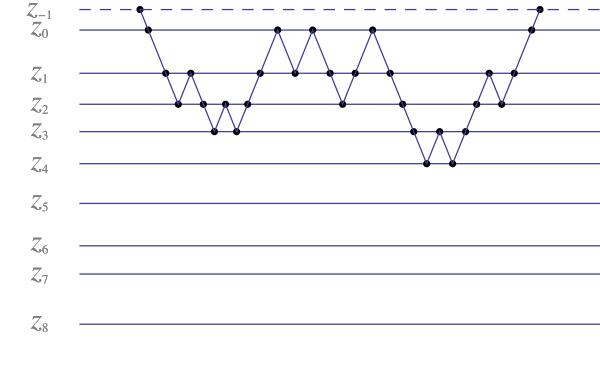}
}
\caption{ The Dyck path for a scattering sequence 
$
\mpp,
$
with the horizontal coordinate now representing time, which increases to the right.}\label{fig-Dyck}
\end{figure}

A scattering sequence $\mpp\in\sss_M$  may be represented graphically as in Figure~\ref{fig-Dyck};  Stanley \cite{St:1999}  calls  such a representation a Dyck path.   
\begin{figure}[h]
\fbox{
\includegraphics[clip,trim=0in .9in 0in 0in, width=4.8in]{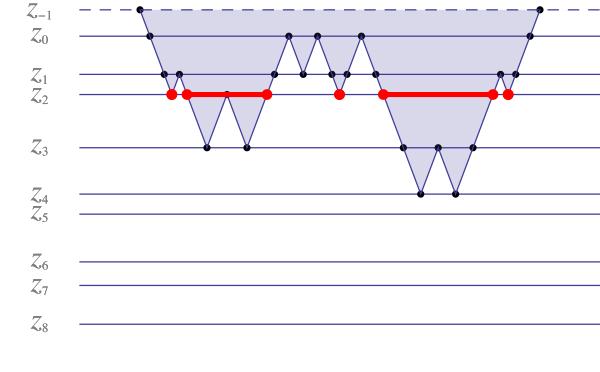}
}
\caption{ The intervals $I^2_j$, in red.  There are five intervals, two of which have positive length, so $k_2=5$ and $b_2=2$.}\label{fig-Intervals}
\end{figure}

Given the Dyck path for a scattering sequence $\mpp\in\sss_M$, let $t$ denote the horizontal coordinate and (as usual) let $z$ denote the vertical coordinate.   Let $U$ denote the portion of the $t,z$-plane on or above the Dyck path and at or below $z=z_{-1}$---the shaded region in Figure~\ref{fig-Intervals}.  For each $n$ in the range $0\leq n\leq M$, consider the horizontal line $L_n$ at depth $z_n$.  The intersection $L_n\cap U$ consists of a disjoint union of closed intervals $I^n_j$, where $1\leq j\leq k_n$; see Figure~\ref{fig-Intervals}.   The intervals of $I^n_j$ are of two types: degenerate intervals consisting of a single points; and non-degenerate intervals having positive length.    Letting $k_n$ denote the total number of intervals $I^n_j$, and letting $b_n$ denote the number of non-degenerate intervals, set 
\[
\kappa(\mpp)=k=(k_0,\ldots,k_M)\quad\mbox{ and }\quad\beta(\mpp)=b=(b_0,\ldots,b_M).
\]

Observe that the entry $k_n$ of the vector $k=\kappa(\mpp)$ counts the number of times the Dyck path crosses back and forth across the $n$th layer $z_{n-1}<z<z_n$; the vector $k$ is therefore called the transit count vector for $\mpp$.   The vector $b=\beta(\mpp)$ is called the branch count vector, for reasons that will be apparent in Section~\ref{sec-tree}.

\subsubsection{The set of all transit count vectors}

The range of $\kappa$ is contained in the set
\begin{equation}\label{LM}
\begin{split}
&\mathfrak{L}_M=\\
&\left\{(k_0,k_1,\ldots,k_M)\in\nat^{M+1}\,\left|\, k_0=1\,\&\,\forall n\leq M-1,\,k_n=0\Rightarrow k_{n+1}=0\right.\right\}.
\end{split}
\end{equation} 
Conversely, for any given $k\in\mathfrak{L}_M$, it is straightforward to construct a realizing scattering sequence.  Thus $\mathfrak{L}_M$ is precisely the range $\kappa(\sss_M)$ of the mapping $\kappa:\sss_M\rightarrow\integer^{M+1}$.

\subsubsection{The formula for arrival times}

The arrival time of $\mpp$, expressed in terms of the transit count vector $k=\kappa(\mpp)$, is simply 
\[
\langle k,\tau\rangle=k_0\tau_0+k_1\tau_1+\cdots+k_M\tau_M.
\]
Therefore $\gtr(t)$ may be written as 
\begin{equation}\label{arrival}
\gtr(t)=\sum_{k\in\mathfrak{L}_M}\mathfrak{a}(R,k)\delta(t-\langle k,\tau\rangle),
\end{equation}
where the amplitudes $\mathfrak{a}(R,k)$ are given by the formula
\begin{equation}\label{aRkdefintion}
\mathfrak{a}(R,k)=\sum_{\substack{\mpp\in\sss_M\; \scriptstyle{\rm such\; that}\\ \kappa(\mpp)=k}}w(\mpp).
\end{equation}
Note that the weight $w(\mpp)$ of a scattering sequence depends only on $R$, and not on $\tau$; the next step is to find an explicit formula for $\mathfrak{a}(R,k)$.   This is greatly facilitated by introducing another representation for scattering sequences, in terms of trees.  

\subsection{The tree representation of a scattering sequence\label{sec-tree}}

\begin{figure}[h]
\parbox{4.1in}{
\fbox{
\includegraphics[clip,trim=.9in .9in 0in 0in, width=3.8in, angle=180]{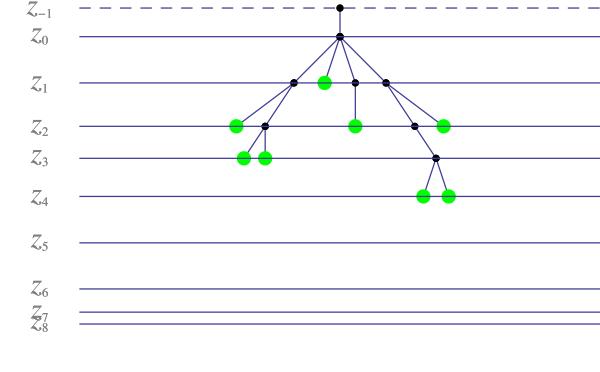}
}
\caption{A tree, with the leaves coloured green. The root is at the bottom, and horizontal lines indicate the various heights of vertices.}\label{fig-aTree}
}
\end{figure}

A tree is a connected cycle-free graph.  The vertices of a tree are divided into three anatomical types, as follows: the root is a single, specially designated vertex; a non-root that belongs to just one edge is called a leaf; all other non-root vertices are called branch points.     Vertices in a tree have a height, determined by their distance (in the sense of shortest path) to the root. See Figure~\ref{fig-aTree}. 

\begin{figure}[h]
\fbox{\parbox{\textwidth}{
(a)\includegraphics[clip,trim=0in .7in 0in 0in, width=2.25in]{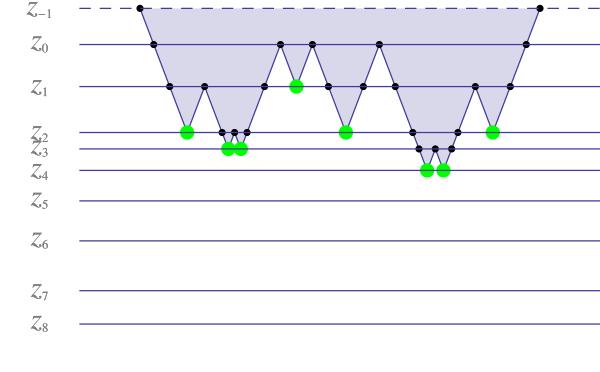}
(b)\includegraphics[clip,trim=0in .7in 0in 0in, width=2.25in]{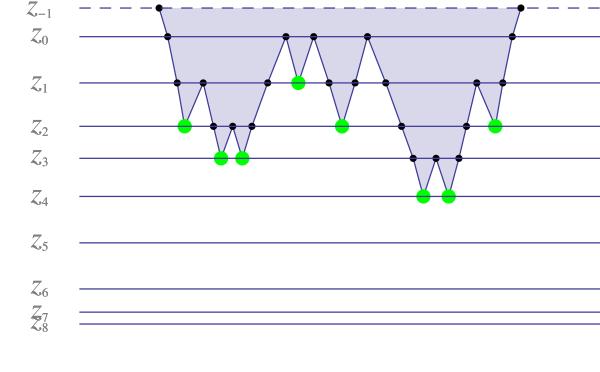}
(c)\includegraphics[clip,trim=0in .7in 0in 0in, width=2.25in]{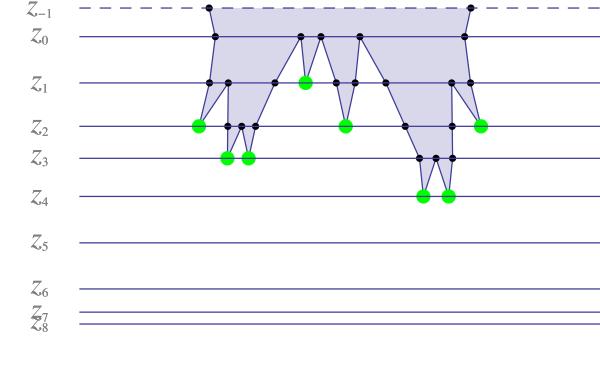}
(d)\includegraphics[clip,trim=0in 0.7in 0in 0in, width=2.25in]{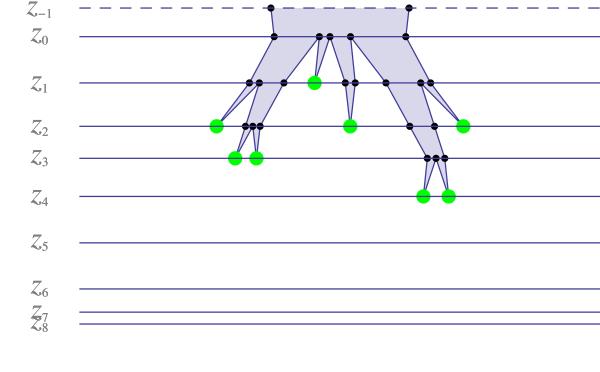}
(e)\includegraphics[clip,trim=0in 0.7in 0in 0in, width=2.25in]{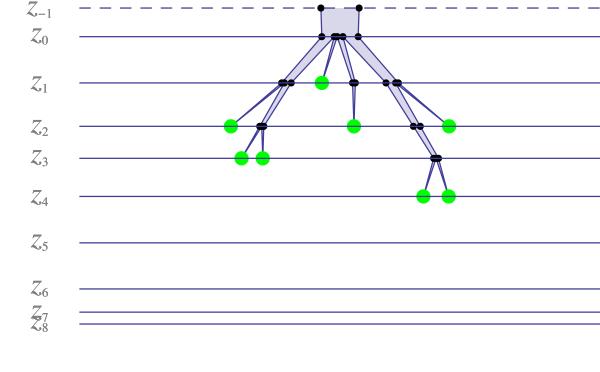}
(f)\includegraphics[clip,trim=0in 0.7in 0in 0in, width=2.25in]{Tree1.jpg}
}
}
\caption{ The collapsing of a Dyck path (a) to a tree (f) (depicted upside down), an operation which is reversible.}\label{fig-tree}
\end{figure}

The association between a scattering sequence and a tree is well-known (see \cite[Exercise 6.19]{St:1999}) and arises, for instance, in the analysis of  Brownian excursions and superprocesses (see \cite[Section~1.1]{Le:2005}).    The tree representing a scattering sequence $\mpp\in\sss_M$ may be obtained simply by collapsing its Dyck path, as follows.   Recall the intervals $I^n_j$ used to define $\kappa(\mpp)$ and $\beta(\mpp)$ above; for present purposes let $I^{-1}_1$ denote the intersection of $z=z_{-1}$ with $U$.  To collapse the Dyck path, contract each of the intervals $I^n_j$ to a point, keeping the distances between intervals unchanged, and interpolate this horizontal contraction linearly on each depth interval $z_{n-1}<z<z_{n}$.  This operation transforms the original Dyck path into a tree  (in fact it is an isotopy between the region $U$ and the resulting tree).  See Figure~\ref{fig-tree}.   Note that the degenerate intervals $I^n_j$ (coloured green for emphasis) end up as leaves, while non-degenerate intervals are contracted to branch points of the tree---except for $I^{-1}_1$, which is contracted to the root.    

Conversely, given a tree, one may recover the original scattering sequence by tracing the outline of the tree, from the root (keeping the tree on the left), and recording the depths of the vertices in the order that they are passed.  

Observe that the vectors $(k,b)=(\kappa(\mpp),\beta(\mpp))$ have a simple interpretation in terms of the tree representing $\mpp\in\sss_M$.   For each $0\leq n\leq M$, $k_n$ is the number of vertices at depth $z_n$, and $b_n$ is the number of branch points at $z_n$.  (This is the reason for calling $b$ the branch count vector for $\mpp$.)   There are some evident constraints on these quantities.  Note first that $b_M=0$.   Furthermore, for $0\leq n\leq M-1$, 
\begin{equation}\label{constraints}
\min\{1,k_{n+1}\}\leq b_n\leq k_{n+1}\quad(0\leq n\leq M-1),
\end{equation}
since each vertex at $z_{n+1}$ is connected by an edge to a unique branch point at $z_n$.    It is convenient to refer to the left shift $\tilde{k}$ of $k=\kappa(\mpp)$; that is,
\begin{equation}\label{ktilde}
\tilde{k}=(k_1,k_2,\ldots,k_M,0)\in\integer^{M+1}.
\end{equation}
In terms of this notation, the constraints
 (\ref{constraints}) become
\begin{equation}\label{constraints2}
\min\{1,\tilde{k}_n\}\leq b_n\leq\min\bigl\{k_n,\tilde{k}_n\bigr\}\quad(0\leq n\leq M).
\end{equation}

An easy application of the tree representation is to determine the possible values of $b=\beta(\mpp)$ for each given $k=\kappa(\mpp)$, or in other words, to determine the set 
\begin{equation}\label{V}
V(k)=\beta(\kappa^{-1}(k)).
\end{equation}
In fact $V(k)$ is determined precisely by (\ref{constraints2}).   
\begin{prop}\label{prop-V}
Given $k\in\mathfrak{L}_M$, 
\[
V(k)=\bigl\{b\,\bigl|\,\min\{\mathbb{1},\tilde{k}\}\leq b\leq\min\{k,\tilde{k}\}\bigr\};
\]
equivalently, $V(k)$ may be expressed as a Cartesian product of sets,
\[
V(k)=V_0\times V_1\times\cdots V_M,\]
where 
$
V_n=\bigl\{\min\{1,\tilde{k}_n\},\min\{1,\tilde{k}_n\}+1,\ldots,\min\{k_n,\tilde{k}_n\}\bigr\}\quad(0\leq n\leq M).
$
\end{prop}
Here $\mathbb{1}\in\integer^{M+1}$ is the vector whose entries are all 1.  The minimum is to be interpreted entrywise, meaning that for $x,y\in\real^{M+1}$, 
\[
\min\{x,y\}=\bigl(\min\{x_0,y_0\},\min\{x_1,y_1\},\ldots,\min\{x_M,y_M\}\bigr)\in\real^{M+1}.
\]
\begin{pf}
The constraints (\ref{constraints2}) imply that $V(k)\subset V_0\times\cdots\times V_M$.   It remains to show that any $b\in V_0\times\cdots\times V_M$ belongs to $V(k)$, which entails showing that there exists a scattering sequence $\mpp$ such that $(\kappa(\mpp),\beta(\mpp))=(k,b)$.   It suffices to construct a tree representing such a $\mpp$, as follows.  Given $b\in V_0\times\cdots\times V_M$, place $k_n$ vertices, consisting of $b_n$ branch points and $k_n-b_n$ leaves (in any order), at depth $z_n$, for $0\leq n\leq M$, and place a root at $z_{-1}$ (considered as a branch point).  Let $N$ denote the largest index such that $k_N\neq 0$.  For $0\leq n\leq N$,
\[
1\leq b_{n-1}\leq\min\{k_{n-1},k_n\},
\]
so there exists a surjection $f_n$ from the $k_n$ vertices at $z_n$ onto the $b_{n-1}$ branch points at $z_{n-1}$.   The tree representing $\mpp$ is completed by drawing an edge from each vertex $v$ at $z_n$ to $f_n(v)$.   
\end{pf}

\subsubsection{A simple formula for the weight}

The tree representation facilitates deriving a simple formula for the weights.  The following lemma uses multi-index notation, whereby given a vector $S=(S_0,S_1,\ldots,S_M)$ and an integer vector $d=(d_0,\ldots,d_M)$,  
\[S^d=\prod_{n=0}^Ms_n^{d_n}.\]  

\begin{lem}\label{lem-weight}
Let $\mpp\in\sss_M$ be a scattering sequence in an $M$-layer model $(\tau,R)$, and set $(k,b)=(\kappa(\mpp),\beta(\mpp))$.  Then
\[
w(\mpp)=(-R)^{\tilde{k}-b}R^{k-b}T^{2b}.
\]
\end{lem}
\begin{pf}
Consider the tree representing $\mpp$.  Observe that each instance of $w_n=R_j$ in (\ref{wn}) corresponds to a unique leaf at $z_j$ (see Figure~\ref{fig-tree}).  Since there are $k_n-b_n$ leaves at $z_n$ this results in a total contribution of $R^{k-b}$.   

Let $v$ be a branch point at $z_j$ having $d_v$ edges to vertices at $z_{j+1}$.   Observe that precisely $d_v-1$ of these edges correspond to an instance of $w_n=-R_j$ in (\ref{wn}), and every occurrence of $w_n=-R_j$ arises this way.  The sum total of numbers $d_v-1$ over branch points $v$ at $z_j$ is simply $k_{n+1}-b_n=\tilde{k}_n-b_n$, making for a total contribution over all depths of $(-R)^{\tilde{k}-b}$.  

Finally, each instance of transmission from the $j$th layer to the $(j+1)$st layer in $\mpp$ corresponds to a vertex in the tree representing $\mpp$ which is not a leaf, i.e., to a branch point at $z_j$; and, since the path $\mpp$ starts and ends at $z_{-1}$ every such transmission has a corresponding return transmission in the opposite direction, from the $(j+1)$st layer to the $j$th layer.  There are $b_j$ branch points at $z_j$, and each of these corresponds to two transmissions across the boundary at $z_j$, making for a total contribution to $w(\mpp)$ of $T^{2b}$.   Since every $w_n$ is covered by one of the above cases, the lemma follows. 
\end{pf}

\subsection{An explicit formula for the impulse response}

A formula for the coefficients $\mathfrak{a}(R,k)$ in (\ref{arrival}), defined as the summation (\ref{aRkdefintion}), is now within easy reach.  Recall that the binomial coefficient $\binom{x}{y}$ for a pair of non-negative integer vectors $x,y\in\integer^{M+1}$, with $y\leq x$, is to be interpreted as 
\[
\binom{x}{y}=\prod_{n=0}^M\binom{x_n}{y_n}.
\]
(The inequality $y\leq x$ means that $x-y$ has non-negative entries.)
\begin{thm}\label{thm-amplitude2} Let $(\tau,R)$ be an $M$-layer model for some integer $M\geq 1$, let $k\in\mathfrak{L}_M$, and set $u=\min\{\mathbb{1},\tilde{k}\}$.  Then
\begin{equation}\label{aRk}
\mathfrak{a}(R,k)=\sum_{b\in V(k)}\binom{k}{b}\binom{\tilde{k}-u}{b-u}(-R)^{\tilde{k}-b}R^{k-b}T^{2b},
\end{equation}
where $V(k)$ denotes the set of $b\in\integer^{M+1}$ such that $u\leq b\leq\min\{k,\tilde{k}\}$.
\end{thm}
\begin{pf}
The total amplitude resulting from scattering sequences having a given transit count vector $k\in\mathfrak{L}_M$ is 
\[
\mathfrak{a}(R,k)=\sum_{\substack{\mpp\in\sss_M\; \scriptstyle{\rm such\; that}\\ \kappa(\mpp)=k}}w(\mpp).
\]
By Proposition~\ref{prop-V} and Lemma~\ref{lem-weight} the above sum may be rearranged as 
\begin{eqnarray*}
\mathfrak{a}(R,k)&=&\sum_{b\in V(k)}\sum_{\substack{\mpp\in\sss_M\; \scriptstyle{\rm such\; that}\\ (\kappa(\mpp),\beta(\mpp))=(k,b)}}w(\mpp)\\
&=&\sum_{b\in V(k)}\Bigl(\#\{\mpp\in\sss_M\,|\,(\kappa(\mpp),\beta(\mpp))=(k,b)\}\Bigr)(-R)^{\tilde{k}-b}R^{k-b}T^{2b}.
\end{eqnarray*}
Thus all that is required is to count the number of scattering sequences having a given transit count vector and branch point count vector.   Equivalently, it suffices to count the number of corresponding trees---which is straightforward.  
Consider first the arrangement of vertices in a tree for which $(\kappa(\mpp),\beta(\mpp))=(k,b)$.   At each depth $z_n$, there are $k_n$ vertices of which $b_n$ are branch points and $k_n-b_n$ are leaves.  There are $\binom{k_n}{b_n}$ ways of arranging these from left to right, making for a total of 
\begin{equation}\label{vertex}
\binom{k}{b}=\prod_{j=0}^M\binom{k_n}{b_n}
\end{equation}
possible vertex arrangements.  (There is only one way to place the root at $z_{-1}$, which may be ignored.)  

For each vertex arrangement there are various possible edge arrangements, as follows.   Each of the $k_{n+1}=\tilde{k}_n$ vertices at $z_{n+1}$ must be connected by an edge to one of the $b_n$ branch points at $z_n$, respecting the vertex ordering (so that edges don't cross).  This is equivalent to choosing a $b_n$-part ordered partition of the integer $\tilde{k}_n$.  If $\tilde{k}_n\geq 1$, there are $\binom{\tilde{k}_n-1}{b_n-1}$ possible choices; and if $\tilde{k}_n=0$ then $b_n=0$ and there is $1=\binom{\tilde{k}_n}{b_n}$ (empty) arrangement.  Letting $N$ denote the largest index for which $\tilde{k}_N\neq 0$, the total number of edge arrangements is 
\begin{equation}\label{edge}
\binom{\tilde{k}-u}{b-u}=\prod_{n=0}^N\binom{\tilde{k}_n-1}{b_n-1}.
\end{equation}
Combining (\ref{vertex}) and (\ref{edge}) yields a total tree count of 
\[
\#\{\mpp\in\sss_M\,|\,(\kappa(\mpp),\beta(\mpp))=\binom{k}{b}\binom{\tilde{k}-u}{b-u},
\]
completing the proof. \end{pf}

Thus the coefficients $\mathfrak{a}(R,k)$ in the expansion (\ref{arrival}) coincide with the amplitude polynomials $a(R,k)$ of Definition~\ref{defn-amplitude-polynomial}, proving Theorem~\ref{thm-amplitude}.

\section{Conclusions\label{sec-conclusion}}

For generic models, Theorems~\ref{thm-model-data} and \ref{thm-data-model} express precisely the local linear-algebraic character of the model-data correspondence, and the inverse algorithm, Algorithm~\ref{alg-inverse}, gives a practical method by which to recover the model from the data.   Figures~\ref{figure-array}, \ref{fig-generic} and \ref{fig-range}  convey the geometric view of the underlying correspondence between travel time vectors $\tau$ and arrival time vectors $\sigma$.   There are several remarks to be made about the implications of this general picture; each of the sections below considers a particular issue.

\subsection{Modifying the boundary depth $z_{-1}$\label{sec-modifying-z}}

The physical setup in which source and receiver sit at depth $z_{-1}$, above the boundary $z_0$ of the layered half-space, may be easily modified to suit a particular application without altering the essential results or even the formulas.   For example, consider the situation of source and receiver at depth $z_{-1}=z_0$.   One has to decide whether a pulse must cross the boundary $z_0$ to be detected by the receiver.   If not, then the formulas in Theorem~\ref{thm-amplitude} remain valid by simply setting $\tau_0=0$ and by replacing each occurrence of the transmission coefficient 
\[
T_0=\sqrt{1-R_0^2}
\]
by 1.  Similar modifications suffice to adapt the results to other physical scenarios that might more accurately model a particular experimental modality, such as marine seismic etc.    The results of the present paper are also relevant to non-seismic applications such as single source-receiver tussue sensing adaptive radar \cite{WiFeWe:2006}, \cite{KuFe:2009}.

\subsection{Non-generic models\label{sec-non-generic}}

What happens as the normalized travel time vector $\hat\tau$ approaches the boundary of one of the cells in Figure~\ref{fig-generic} or its higher dimensional analogues?    There are two subcases to consider.   If $\hat\tau$ crosses one of the orange lines---so that the corresponding lattice set $\LtM$ remains constant---then the enumeration function permutes the values of two $k,k^\prime\in\LtM$.   That is, there is an $n$ such that on one side of the line $\psi(k)=n$ and $\psi(k^\prime)=n+1$, while on the other side $\psi(k^\prime)=n$ and $\psi(k)=n+1$, this being the only change in $\psi$.   On the line itself $\psi$ ceases to be injective, mapping $k$ and $k^\prime$ to the same value.   The second case is where $\hat\tau$ crosses a black line.  In this case the lattice set $\LtM$ gains or loses an element (so that the dimension of $A_\psi$ goes up or down accordingly).   On the black line there is a transit count vector $k\neq k^M$ such that $\psi(k)=\psi(k^M)$.    The data corresponding to non-generic model associated with $\hat\tau$ on one of the lines need not determine a unique model, as in Theorem~\ref{thm-ill-posed}.   But this does not mean that the inverse problem cannot be dealt with.   (After all, a number of historical treatments are restricted to the case of equal travel times, a particular non-generic scenario.)   If an extra quantity, $t_{\max}=|\tau|$, is included as part of the data, then the model can be recovered.   However, there is a qualitative difference: the travel time inversion and amplitude inversion no longer fully decouple as they do in Alogorithm~\ref{alg-inverse}.   Instead, the two must be interwoven, resulting in a slower algorithm that has additional steps.   Thus the generic case is not only typical, it is also qualitatively simpler than the non-generic case.

\subsection{Coarsely layered media\label{sec-coarsely-layered}}

In working with finite precision, the meaning of generic must be appropriately adapted.   Referring again to Figure~\ref{fig-generic}, the normalized travel time vector $\hat\tau$ must be sufficiently far from the set of line segments to be considered as effectively generic.   This rules out the regions of the triangle close to the upper boundary where the lines are clustered ever more densely.   Roughly speaking, there is a remaining part of the triangle that corresponds to effecively generic travel times---interior to the larger cells and sufficiently far from the boundary.   A similar restricted set of reflection vectors $R$, sufficiently far from the algebraic hypersurfaces on which $a(R,k)=0$ (for appropriate $k$ as in Section~\ref{sec-generic-definition}), completes the description of effectively generic models.   It seems appropriate to refer to the layered media corresponding to such generic models as being coarsely layered.   Not only is the distribution of depths restricted for such models, so is the number of layers.   (This is because the relative volume of cells in the higher dimensional analogues of Figure~\ref{fig-generic} goes down with increasing dimension.)    Without attempting to quantify the notion more precisely, there is a regime of coarsely layered media in which the deterministic approach of the present paper constitues a practical theory.   For finely layered media where there is no effectively generic model, the stochastic methods of \cite{FoGaPaSo:2007} are more appropriate.   In any case the results presented here provide tools---both in the form of explicit formulas and, more broadly, a geometric perspective---with which to assess the utility and limitations of a deterministic approach to the 1-D acoustic reflection problem.

\bibliography{ReferencesEAL}

\end{document}